\documentclass[manuscript]{aastex}

\shorttitle{vela X-1}
\shortauthors{Maitra \& Paul}

\begin{document}

\title{Pulse phase resolved spectroscopy of Vela X-1 with Suzaku}

\author{Chandreyee Maitra\altaffilmark{1} and Biswajit Paul}
\affil{Raman Research Institute, Sadashivnagar, Bangalore-560080, India }
\email{cmaitra@rri.res.in; bpaul@rri.res.in}
\altaffiltext{1}{Joint Astronomy Programme, Indian Institute of Science, Bangalore-560012, India}

\begin{abstract}
 We present a detailed pulse phase resolved spectral analysis of the persistent high mass X-ray binary pulsar Vela X-1 observed
with \emph{Suzaku} during June $2008$. The pulse profiles exhibit both intensity and energy dependence with multiple peaks at
low energies and double peaks at higher energies. The source shows some spectral evolution over the
duration of the observation and care has been taken to average over data with minimum spectral variability
 for the analysis. We model the continuum with a phenomenological partial
covering high energy cutoff model and a more physical partial covering thermal comptonization
model (CompTT) excluding the time ranges having variable hardness ratio and intensity dependence. For both the models we detect a cyclotron resonant scattering feature (CRSF)
and its harmonic at $\sim 25$ keV and $\sim 50$ keV. Both the CRSF fundamental 
and harmonics parameters are strongly variable over the pulse phase, with the ratio of the two line
energies deviating from the classical value of $2$. The continuum parameters also show significant 
variation over the pulse phase
and give us some idea about the changing physical conditions that is seen with the changing viewing angle at different pulse phases
and obscuration by the accretion stream at some pulse phases.
\end{abstract}


\section{Introduction}

Vela X-1 (4U 0900-40) is a an eclipsing and persistently active high mass X-ray binary (HMXB) system
consisting of a massive ($23 M_{\sun}$; $34 R_{\sun}$) B0.5 1b supergiant HD 77581 \citep{brucato1972,nagase1989};
and a $\sim 1.77  M_{\sun}$ neutron star \citep{rawls2011}. It has an orbital period
of 8.964 days \citep{vankerkwijk1995} with a slightly eccentric orbit (e $\sim$ 0.1). Being a closely spaced binary system \citep{quaintrell2003},
the neutron star is deeply embedded in the strong stellar wind of the companion which has a mass loss rate in excess of $ 10^{-7}$ $M_{\sun} yr^{-1}$
\citep{nagase1986,sako1999}. It is a prototype candidate for accretion via stellar wind, and the typical X-ray luminosity
is of the order of $\sim$ $4*10^{36}$ ergs $s^{-1}$. It exhibits pulsations with a period $\sim$ 283 s. The source is extremely variable undergoing
 giant flares (where flux increases by a factor of $20$), and abrupt off-states, where
the source is almost undetectable for several pulse periods \citep{inoue1984,kreykenbohm2008,doroshenko2011}.\\

The structure of the wind and hence the flaring behavior of Vela X-1 was investigated by \citet{furst2010}
who also found a systematic trend in the change of column density of absorbing matter along the line of sight 
in different phase of the the orbit.
The pulse profiles of Vela X-1 show a complex structure with multiple energy dependent dips. This has been explored 
before for other sources for example in GX 1+4 \citep{galloway2001}, RX J0812.4-3114 \citep{corbet2000}, 1A 1118-61 \citep{devasia2011}, GX 304-1 \citep{devasia2011a} with \emph{RXTE},
 KS 1947+300 \citep{naik2006} with \emph{Beppo-SAX}, GRO J1008-57
 \citep{naik2011}, and 1A 1118-61 \citep{maitra2011} with \emph{Suzaku}.
Dips in the pulse profile can be naturally explained 
due to the absorption in the accretion stream that is phase locked with the neutron star.
The X-ray continuum spectrum of Vela X-1 has been modeled with a power law with an exponential cutoff \citep{white1983,tanaka1986},
or with a Negative Positive EXponential model (NPEX) \citep{orlandini1998,kreykenbohm1999,kreykenbohm2002}.
The continuum is further modified by a considerable amount of photoelectric absorption at low energies which varies 
erratically on short timescales and also with the orbital phase \citep{haberl1990}. A strong iron fluorescence feature
at $\sim$ 6.4 keV has also been reported \citep{watanabe2006}.\\
An absorption like feature at $\sim$ 55 keV was first reported by \citet{kendziorra1992} using data from \emph{HEXE}.
This was interpreted as a Cyclotron Resonance Scattering Feature (CRSF). CRSF arises due to resonant scattering of electrons
in Landau levels in the presence of high magnetic fields. The energy of the CRSF is given by
 $E_{c}=11.6 keV\times\frac{1}{1+z}\times\frac{B}{10^{12}G}$, where $E_{c}$ is the centroid energy, z is the gravitational redshift
and B is the magnetic field strength of the neutron star. CRSF therefore gives us an estimate of the  magnetic field strength
of the neutron star. Pulse phase resolved spectroscopy of the CRSF parameters can provide crucial information on the
magnetic field geometry and the emission region of the neutron star as we probe it at different viewing angles with spin phase.
\citet{makishima1992} and \citet{choi1996} reported
an absorption feature at $\sim$ 32 keV from the Ginga data. \citet{mihara1995} performed a detailed analysis of the Ginga observations
and found that the CRSFs vary in strength over the pulse phase. 
\citet{kretschmar1997} also confirmed the existence of two absorption features at $\sim$ 23 keV and $\sim$
45 keV using the broadband data from the HEXE and TTM, and found similar variations with pulse phase as from the Ginga data.
Subsequently more detailed pulse phase resolved analysis was performed by \cite{kreykenbohm1999,kreykenbohm2002} using RXTE
 data who confirmed the existence of the two lines at $\sim$ 25 \&
55 keV and also reported 
it's variation with the pulse phase. \cite{orlandini1998} and \cite{labarbera2003} performed phase resolved analysis of Vela X-1
using \emph{BeppoSAX} data but could only confirm the existence of the absorption line at $\sim$ 55 keV.\\

We present here a pulse phase resolved spectral analysis of Vela X-1 using \emph{Suzaku} observation in June 2008.
The same \emph{Suzaku} observation has been analyzed by \cite{doroshenko2011} who have mainly focused on the
three off-states detected for a few pulses during the observation, and have detected pulsations for the first time 
in these ``off-states``. They have also analyzed the energy dependent pulse profiles and the energy spectrum in
both the off and normal states, and have confirmed the presence of the two CRSFs in the normal states. We have however
probed the energy dependence of the pulse profiles in much more detail and carried out pulse phase resolved spectroscopy
in narrow phase bins to explore its complex nature.
The variations of the continuum spectral parameters as well as the CRSF fundamental and harmonic with pulse phase reported here are the 
most detailed results available so far. While performing pulse phase resolved analysis, care has also been taken to average over
data having similar intensity and spectral hardness, to probe the variation solely due to the changing viewing angle with
pulse phase. In section $2$ we describe the observation and data reduction, in section $3$ the pulse profiles including
it's energy and intensity dependence, in section $4$ spectroscopy, including phase averaged and phase resolved
followed by discussions and conclusions in section $5$. 

\section{Observations \& Data Reduction}
Vela X-1 has been observed with \emph{Suzaku} \citep{mitsuda2007} on June 17-18 2008 (Obs. Id--403045010). The observation
 (exposure $\sim$ 103 ks) was performed at 'HXD nominal' pointing position. \emph{Suzaku} has two sets of instruments
on board, the X-ray Imaging Spectrometer XIS \citep {koyama2007} covering the 0.2-12 
keV range, the Hard X-ray Detector (HXD) having PIN diodes \citep {takahashi2007} 
covering the energy range of 10--70 keV, and GSO crystal scintillator detectors covering the 70--600 keV
energy band. The XIS consists of four CCD detectors, three of which are front illuminated (FI :
XIS0, XIS2, XIS3) and one back 
illuminated (BI: XIS1). Since the loss of XIS2 in 2006 November, only XIS 0, 1 \& 3 are operational.
For this observation, the XIS's were operated in 'Normal' clock mode 
in the '$\frac{1}{4}$ window' option which gave a time resolution of 2 s. \\
HEASOFT 6.11 was used for the reduction and extraction. 
For the reduction and extraction of the XIS data,
 the unfiltered XIS events were reprocessed with the CALDB version 20100123 for XIS and 20091007 for HXD.
The reprocessed XIS event files were then checked for the possible presence of photon pile-up. 
Pile-up estimation was performed by examining the Point Spread Function (PSF) 
of the XIS's by checking the count rate per one CCD exposure at the image peak as given 
by Yamada \& Takahashi.  \footnote{http://www-utheal.phys.s.u-tokyo.ac.jp/~yuasa/wiki/index.php\\/How\_to\_check\_pile\_up\_of\_Suzaku\_XIS\_data}
Following their procedure, the value obtained was higher than that of Crab Nebula count rate of 36 ct/sq arcmin/s (CCD exposure), and the radius
at which this value equals 36 in the PSF is about 15-16 arcsec. Thus the check performed showed a case of moderate photon
pile up, and 15 pixels from the image center were removed to account for this effect.
 For the extraction of XIS light curves and spectra from the
reprocessed XIS data, a circular region of $4^{'}$ diameter
was selected around the source centroid with the central 15 ($\sim$ $16^{''}$) pixels removed.
 Background light curves and spectra were extracted by selecting a region of the same size away from the source.
The XIS count rate after the pile-up correction was around $\sim$ $34$  c $s^{-1}$ with $\sim 12 \%$ of the 
photons being lost due to the removal of the photons from the central region.  Response files and effective area files were generated by using the 
FTOOLS task 'xisresp'. For extracting  HXD/PIN light curves and spectra, cleaned event files were used (processing version 2.2.8.20).
For HXD/PIN background, simulated 'tuned' non X-ray background event files 
(NXB) corresponding to June 2008 were used to estimate the non X-ray 
background\footnote{http://heasarc.nasa.gov/docs/suzaku/analysis/pinbgd.html}, and the cosmic X-ray 
background was simulated as suggested by the instrument team\footnote{http://heasarc.nasa.gov/docs/suzaku/analysis/pin\_cxb.html} applying appropriate 
normalizations for both cases. Response files of the respective observations were obtained from 
the \emph{Suzaku} guest observatory facility.\footnote{http://heasarc.nasa.gov/docs/heasarc/caldb/suzaku/ and used for the HXD/PIN spectrum}

\section{Timing Analysis}

\begin{figure}

\begin{center}
\includegraphics[scale=0.7]{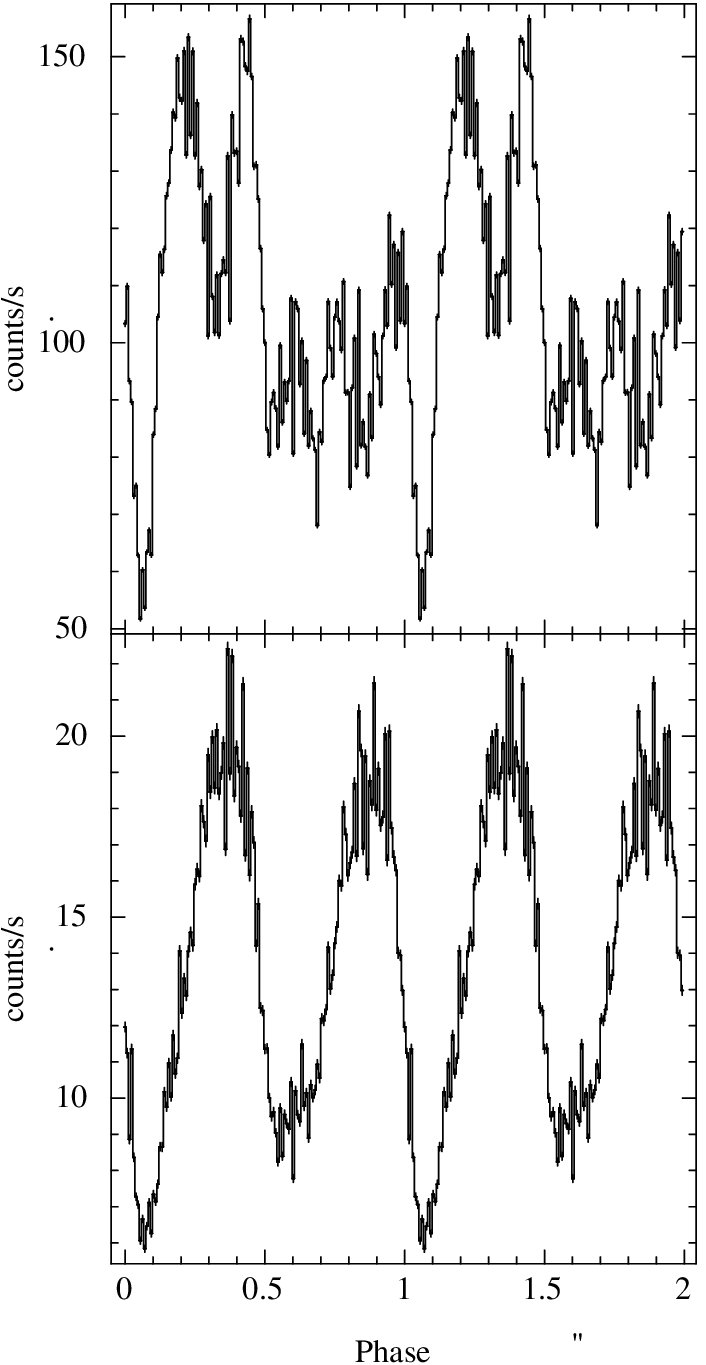}
\end{center}
\caption{Average pulse profile of Vela X-1 folded with the best obtained period.
The top panel shows the pulse profiles
(0.3-12 keV) using XIS data \& the bottom panel shows the pulse profile (10-70 keV) using PIN data. }
\label{fig1}
\end{figure}

For the timing analysis, we applied barycentric corrections to the event data files using the FTOOLS 
task 'aebarycen'. Light curves with a time resolution of 2 s and 1 s were extracted from the XISs (0.2--12 keV) 
and the HXD/PIN (10--70 keV), respectively. Pulsations were searched in the data by applying pulse folding and $\chi^{2}$ maximization technique. 
The best estimate of the period was found to be $283.24 \pm 0.16$ s. 
We also checked for the correction of the pulse arrival times due to 
the orbital motion of the neutron star which showed that the filtered events corresponded to an orbital phase $\sim$ 0.17 to 0.36. The orbital motion 
curve however fitted well to a straight line in this phase range. This implies only a linear correction in the pulse arrival times which
would not alter our results on pulse phase resolved spectroscopy
significantly. Thus orbital correction for the pulse arrival times was not
required.
 The pulse period has been determined for the same dataset to a greater precision
by \cite{doroshenko2011}. Figure~\ref{fig1} shows the average pulse profile of Vela X-1 created by folding the light curve with the
obtained pulse period.

\subsection{Light curves \& hardness ratio}
\begin{figure}
\begin{center}
\includegraphics[scale=0.5,angle=-90]{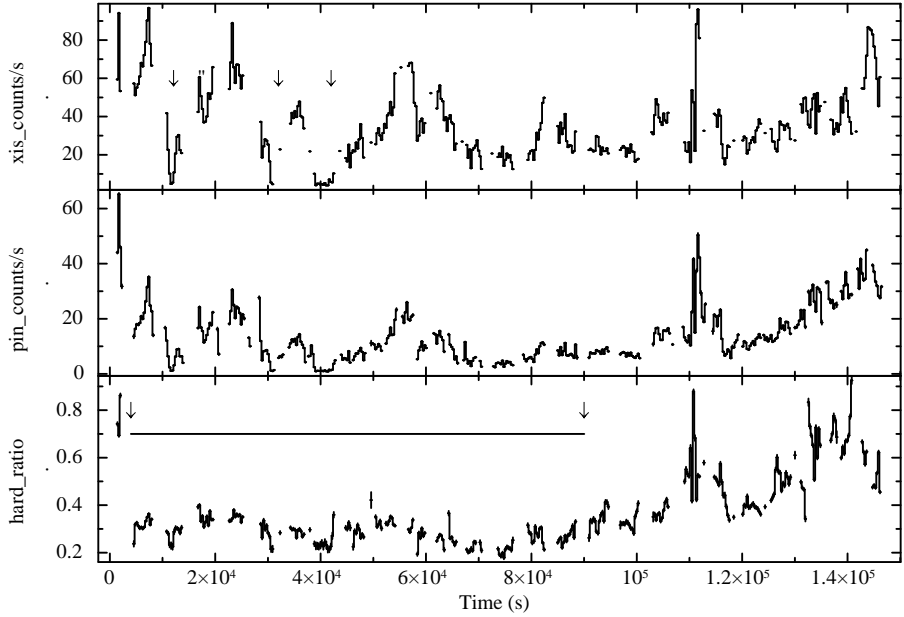}
\end{center}
\caption{The first panel shows the light curve of Vela X-1 obtained with 
\emph{Suzaku} for one of the XIS in the energy band of 0.3-12 keV. The second panel shows the same obtained
in the PIN energy band(10-70 keV). The time binning is equal to the pulse period. The bottom panel shows the hardness ratio. The arrows and the
dash indicates the time range of the data which were selected on the basis of constant hardness ratio 
for further analysis.}
\label{fig2}
\end{figure}

Figure~\ref{fig2} shows the light curves of Vela X-1 rebinned with the timebin equal to the pulse period as obtained from the XIS and PIN data respectively. The light curves
show that the source is highly variable. Panel 3 of the same figure shows the hardness ratio (ratio of PIN counts to XIS counts)
for the entire stretch. Except for the brief initial stretch at the beginning, the hardness ratio remains nearly constant upto the first 90000 s after which it is more variable with 
an increasing trend. For this above mentioned constant hardness ratio stretch, arrows in panel 1 also show the
"off-states" in the XIS light curve as is presented in \cite{doroshenko2011}. 
Since the main aim of this work is to perform pulse phase resolved spectroscopy to probe the spectral parameters with
changing viewing angle, we took only this stretch of constant
hardness ratio excluding the ``off-states`` for further analysis. This would ensure that our results would be free from any systematic effects of phase average spectral variability.

\subsection{Intensity dependence of the pulse profiles}
As already mentioned earlier, since the main aim of this work is to probe the pulse phase dependence of the spectral parameters,
we also checked for the intensity dependence of the pulse profiles to avoid averaging over data having different luminosity dependence.
Intensity dependent pulse profiles were created using both XIS (0.3-12 keV) and PIN (10-70 keV) for the above mentioned
stretch of the data having similar hardness ratio and excluding the ''off-states". Pulse profiles were created in three intensity windows having characteristic XIS count rates
$\le$ 20 c $s^{-1}$, 20-50 c $s^{-1}$ and
$\ge$ 50 c $s^{-1}$ by applying intensity filtering. Figure 3 shows the intensity dependent pulse 
profiles in the XIS and PIN energy band respectively. The XIS and the PIN profiles also show similar 
characteristics in all the intensity bands except for a slightly larger pulse fraction in XIS
at low count rate ($\le$ 20) and in PIN at medium count rate (20-50).
 The XIS pulse profile created with counts
greater than 50 however differs considerably, with the dips and peaks in some phases not coinciding as can be seen from the figure, the dip in the
highest intensity profile at phase $\sim$ 0.6 coincides with a peak in the other intensity band profiles. To avoid this intensity
dependence we averaged over data only in the intensity band of $\le$ 50 c $s^{-1}$ to perform spectral analysis.

\subsection{Energy dependence of the pulse profiles}
Energy resolved pulse profiles were created by folding the light curves in different energy bands with the obtained pulse period . In light of the issues discussed in the above sections, pulse profiles were created for the
constant hardness ratio stretch excluding the "off-states" and in the intensity band of $\le$ 50 $s^{-1}$. This is different from that presented by \citet{doroshenko2011} who have
compared the energy resolved pulse profiles between the on and ``off-states''. Our main goal is to avoid the systematic effects which might affect our results of
pulse phase resolved spectroscopy described later in section 4.2.
The energy dependence of the pulse profiles is shown in  Figure~\ref{fig4}. The pulse profiles from 0.3-12 keV are created using all the 
three XISs (0, 1 \& 3). Higher energy profiles are created from the PIN data. The figure shows that the pulse profiles have complex energy dependent structures, 
with upto five peaks in the XIS 
energy band which merge to two asymmetric peaks in the PIN energy band. With a careful examination of the pulse profiles,
the following characteristics can be observed (refer Figure~\ref{fig4}
for the exact location of the low energy peaks)
\begin{enumerate}
\item The first two peaks in the low energy profiles at phases $\sim$ 0.2 and 0.5 (henceforth termed peaks a and b) merge
to become a single asymmetric peak (phase $\sim$ 0.3) at $\ge$ 12 keV. The next three peaks of the low energy profiles at phases $\sim$ 0.6, 0.8 and 1 (henceforth termed peaks
c, d and e) merge to become the second high energy peak (phase $\sim$ 0.9) after 12 keV.
\item The dip at phase $\sim$ 0.3 between the peaks a and b decreases with energy and disappears at $\sim$ 12 keV.
\item The dip at phase $\sim$ 0.7 between the peaks c and d however increases with energy upto 5 keV after which it shows
 a decreasing trend and disappears at $\sim$ 14 keV.
\item The dip at phase $\sim$ 0.9 between the peaks d and e also decreases with energy and disappears at $\sim$ 7 keV.
\item There is also an indication of a dip like structure at the lowest energies ($\le$ 5 keV) between phase 0.5-0.6.
\end{enumerate}

\section{Spectroscopy}
\subsection{Pulse phase averaged spectroscopy}
 We performed pulse phase averaged spectral analysis 
of Vela X-1 using spectra from the front illuminated CCDs spectrum (XISs-0 and 3), back illuminated CCD spectrum (XIS 1) and the
 PIN spectrum. As discussed earlier we extracted spectra only for the above mentioned stretch of the data having similar hardness ratio, excluded the "off-states",
and further applied intensity filtering to choose the time within it having count rates $\le$ 50 c $s^{-1}$. 
Spectral fitting was performed using \emph{XSPEC} v12.7.0. The energy ranges chosen for the fits were 0.8-10 keV 
for the XISs and 12-70 keV for the PIN spectrum 
respectively. Due to an artificial structure in the XIS spectra around the Si edge and Au edge, the  energy range of 1.75-2.23 keV was  
neglected. After appropriate background subtraction we fitted the spectra simultaneously with all parameters tied, except the relative instrument normalizations 
which were kept free. The XIS spectra were rebinned by a factor of 6 from 0.8-6 keV and 7-10 keV, and by
a factor of 2 between 6-7 keV. The PIN spectra were rebinned by a factor of
 2 upto 22 keV, by 4 upto 45 keV, and 6 upto 70 keV. \\ The continuum was fitted with both a phenomenological high-energy cutoff 'highecut'
model and a more physical comptonization model 'CompTT' assuming a spherical geometry for the Comptonizing region \citep{titarchuk1994}. In both the cases  along with the
Galactic line of sight absorption, a partial covering absorption model 'pcfabs' was applied to take into account the intrinsic absorption local to the neutron star.
 A gaussian line was also used to model the narrow Fe k$\alpha$ feature found at 6.4 keV. \newline In addition, two absorption features were found at 25 and
50 keV, which are the CRSF fundamental and harmonic as found previously for this source. These were modeled with Lorentzian profiles. Figure~\ref{fig5} shows 
the best-fitting spectra along with the residuals. 
Since the energy spectrum has a strong pulse phase dependence, we do not expect the pulse
phase averaged spectrum to fit well to a simple continuum model. The best-fitting parameters for the
high energy cutoff 'highecut' model and the 'CompTT' model are given in Table 1. The best fit had a 
reduced $\chi^{2}$ of 2.03 for 831 d.o.f for the 'highecut' model and 1.92 for 832 d.o.f for the 'CompTT' model
respectively. The value of the additional absorption component (\ensuremath{N_{\mathrm{H2}}}) and the covering fraction (\ensuremath{Cv_{\mathrm{fract}}}) are higher
for the 'CompTT' model. The fundamental CRSF has similar parameters for
both the models except for a slightly wider line in the case of the 'CompTT' model.
The harmonic is however much deeper and wider in the case of 'CompTT' model. 
The continuum model used by \citet{doroshenko2011} is slightly different from ours since they used
a combination of powerlaw and CompTT model while we have used a partial covering CompTT
model. As a result, the Galactic line of sight absorption (\ensuremath{N_{\mathrm{H1}}}) value obtained in
our analysis is lower, and the comptonization parameters are also
different. The cyclotron parameters however match well and are consistent with each other within
error bars. Keeping in mind that Ecycl from
the Lorentzian description does not define the minimum of the line profile \citep{nakajima2010}
, from the values of \ensuremath{E1_{\mathrm{cycl}}}
and \ensuremath{E2_{\mathrm{cycl}}} we obtained with the Lorentzian (cyclabs) model, the
corresponding minimum of the line profile are calculated to be at
26.8 keV and 56.3 keV which are
consistent with the values obtained by \citet{doroshenko2011} within error bars. The depth and width
of the cyclotron lines are also consistent.\\

Confidence contours were plotted to check the inter dependence between some of the parameters
for both the models.  Figure \ref{fig11} shows the $\chi^{2}$ confidence contours between
several pairs of model parameters for the phase averaged spectra. Among the continuum parameters when both were allowed to vary, the partial covering
parameters \ensuremath{N_{\mathrm{H2}}} and \ensuremath{Cv_{\mathrm{fract}}} could be constrained well within 14 \% and 8 \% of the respective parameter values
for the 'highecut' model. There were however limitations in this model like the break energy 'E-cut'
and the fundamental CRSF \ensuremath{E1_{\mathrm{cycl}}} were found to be correlated and could not be well constrained independently.
The contour between \ensuremath{E1_{\mathrm{cycl}}} and the powerlaw photon index ($\Gamma$) showed that although $\Gamma$ could
be constrained very well, \ensuremath{E1_{\mathrm{cycl}}} could not be.
The fundamental being very shallow, and possibly because of the correlation with the break energy and 
inability to constrain against
$\Gamma$
in the 'highecut'
model, the energy and depth of the fundamental could not be constrained completely independent 
of these parameters. However, similar trend of the variation of the fundamental line energy
obtained with pulse phase for both the models give us an added confidence
on our results. \ensuremath{N_{\mathrm{H2}}} and \ensuremath{Cv_{\mathrm{fract}}} for the 'CompTT'
model also could be constrained well (11 \% and 6 \% of their values). The optical depth
and electron temperature '$\tau$' and '$kT$' showed some correlation
but could be constrained between 18 \% and 4 \% of their values. The interdependence between the parameters discussed above
were also evident from the results
of the pulse phase resolved spectroscopy of the parameters discussed later. For the cyclotron line parameters,
both the CRSFs could be constrained well in the case of the 'CompTT' model and showed no dependence both among themselves
or with the continuum parameters like '$\tau$' and '$kT$'.
 In the case of 'highecut' model although parameters of the CRSF harmonic
could be constrained well, the \ensuremath{E1_{\mathrm{cycl}}} and \ensuremath{D1_{\mathrm{cycl}}} contour maps showed
dependency and difficulty in constraining both the parameters simultaneously. However the similar trend of variation of the parameters
obtained with both the spectral models as discussed above gives us confidence on our obtained results.
Because of some of the limitations
discussed here, the errors quoted and plotted henceforth are assuming variation of only one parameter, with the other parameters allowed to vary independently to find the
minimum $\chi^{2}$.

\subsection{Pulse phase resolved spectroscopy}

Strong energy dependence of the pulse profiles, as shown in Figure 4 implies dependence of the energy spectrum on the pulse phase. 
The narrow energy dependent dips in the pulse profiles can also be explained by a partial covering absorption model in which 
the absorber is phase locked with the neutron star.
To investigate this we performed pulse phase resolved spectroscopy, applying phase filtering in the
FTOOLS task XSELECT. For the phase resolved analysis we used the same background spectra and response matrices as was used for the phase
averaged spectrum for both the XIS's and the PIN data. We also fitted the spectra in the same energy range and rebinned them by the same factor
as in phase averaged case. The value of the Galactic absorption (\ensuremath{N_{\mathrm{H1}}}) and the Fe line width were frozen to the 
phase averaged values for the two respective models. \\ 
\subsubsection{Phase resolved spectroscopy of the cyclotron parameters} 
To investigate the pulse phase resolved spectroscopy of the two CRSFs we generated phase resolved spectra with the phases centered 
around as in the case of 25 independent bins but at thrice their widths. This resulted in 25 overlapping bins out of which
only 8 were independent. Due to limited statistics we were unable to constrain all the parameters, and
froze the width of the CRSF harmonic to the phase averaged value of the respective models
and varied the rest of the parameters with pulse phase. For phases close to or at the off pulse regions,
the CRSFs are weak and sometimes cannot be detected in the spectra for either or both of the continuum models.
Confidence contours were created for the pairs of parameters as shown in panels 'B', 'C', 'D', 'E', 'H', 'I', 'J', 'K' in Figure 6 for some of the phase resolved spectra.
The fundamental  \ensuremath{E1_{\mathrm{cycl}}}, when paired with the parameters \ensuremath{D1_{\mathrm{cycl}}}, E-cut or $\Gamma$ could not be constrained simultaneously as was the case with the phase averaged spectrum.
 For all the other pairs, the parameters could be constrained fairly well with the errors being $\sim$ 1.5 times the corresponding errors with the phase averaged spectrum.
Refer section 4.1

Figure \ref{fig6} shows the variation of the cyclotron parameters for both the 
models as a function of pulse phase. 
Despite statistical limitations in the data to constrain the CRSF fundamental energy and depth
independent of the cutoff energy in the 'highecut' model, the variation of the CRSF parameters have
very similar pattern for both the 'highecut' and the 'CompTT'model. This gives us reasonable amount
of confidence on the obtained results. The results also follow the trend as
obtained previously by \citet{kreykenbohm2002} using \emph{RXTE} but they did not have enough sensitivity to probe these features in detail.
The following features can be seen from the Figure \ref{fig6}. The variation is compared with respect to the
high energy PIN pulse profile shown in panel two of the same figure and the peaks and the dips refer to the peaks
and dips in the same.
\begin{enumerate}
 \item The energy of the fundamental (\ensuremath{E1_{\mathrm{cycl}}}) vary by a large amount with pulse phase, with the values varying
between 22 keV in the off pulse region to 28 keV in the ascending phase of the second pulse. It  has a decreasing trend with phase for the first pulse with nearly constant values for the second pulse, except for the ascent of the second pulse
at phase $\sim$ 0.7 where E1 peaks in value.
\item The depth of the fundamental (\ensuremath{D1_{\mathrm{cycl}}}) varies almost by a factor of three with the pulse phase, with the deepest lines near the pulse peaks and shallowest in the off pulse regions. 
 Its value peaks near the ascending edges of the first pulse (phase $\sim$ 0.2-0.4) and the ascending edge of the
second pulse (phase $\sim$ 0.7). The lines are in general deeper for the first pulse.
\item The width (\ensuremath{W1_{\mathrm{cycl}}}) of the fundamental varies almost by a factor of two. The width is maximum at the rising edges of the two pulse peaks at phases
$\sim$ 0.1 and 0.7.   
\item The energy of the second harmonic (\ensuremath{E2_{\mathrm{cycl}}}) varies between values of 48 keV to 62 keV. The values are maximum at the descending phase of both the pulses.
\item The depth of the second harmonic (\ensuremath{D2_{\mathrm{cycl}}}) varies by a factor of 2 with the deepest harmonics found at the same phases
as the phases having maximum \ensuremath{E2_{\mathrm{cycl}}}.
\item The ratio of the two line energies \ensuremath{E1_{\mathrm{cycl}}}/\ensuremath{E2_{\mathrm{cycl}}} has significant pulse phase dependence (Figure \ref{fig9}) 
with an average value of $\sim$ 2.1, maximum of $\sim$ 2.6 near the pulse peaks at phase $\sim$ 0.4 and $\sim$
0.9 
and minimum of $\sim$ 1.7 near the
off pulse regions at phase $\sim$ 0.8.
\end{enumerate}
\subsubsection{Phase resolved spectroscopy of the continuum parameters}
To investigate the pulse phase resolved spectroscopy of the continuum parameters we generated the phase resolved spectra with 25 independent phase
bins. We froze the cyclotron parameters of the corresponding phase bins to the best-fit values obtained for the investigation of the cyclotron
line parameters using 25 overlapping phase bins as described in section 4.2.1. Figure  \ref{fig7} shows the best-fit phase resolved continuum parameters for both the models as
a function of the pulse phase. Confidence contours were created for the pairs of parameters as shown in panels 'A', 'F' and 'G' in Figure 6 for some of the phase resolved spectra.
 For all the pairs, the parameters could be constrained well with the errors being $\sim$ 1.3 times the corresponding errors with the phase averaged spectrum.
Refer section 4.1
The results obtained as seen from the figure from both the models are as follows:
\begin{enumerate}
 \item At the dips between peak a and b (phase $\sim$ 0.3) and peak d and e (phase $\sim$ 0.9) of the low energy XIS profile, there is a sudden increase in the value of 
the local absorption component (\ensuremath{N_{\mathrm{H2}}}). The dip being larger at phase 0.3 the corresponding value of \ensuremath{N_{\mathrm{H2}}} is also larger.  
\item At the main dip of the XIS profile between phase 0.5-0.6, there is also an indication of an increase of \ensuremath{N_{\mathrm{H2}}} with a high covering fraction (\ensuremath{Cv_{\mathrm{fract}}}).  
\item The dip between c and d (phase $\sim$ 0.7) of the XIS profile has somewhat different characteristics and extends upto 20 keV. The 'highecut' model
shows a very high value  \ensuremath{N_{\mathrm{H2}}} and a very small \ensuremath{Cv_{\mathrm{fract}}} at this phase. The 'CompTT' model however
does not show any significant change in the values of \ensuremath{N_{\mathrm{H2}}} or  \ensuremath{Cv_{\mathrm{fract}}}, but a very high optical depth ($\tau$).
\item The value of \ensuremath{Cv_{\mathrm{fract}}} shows somewhat different pattern of variation for the two models. For the 'highecut' model, the phases where \ensuremath{N_{\mathrm{H2}}} is high usually has a small \ensuremath{Cv_{\mathrm{fract}}},
but for the 'CompTT' model there is an indication of correlation between the two parameters.
\item In general the optical depth ($\tau$) increases and the low-energy seed temperature \ensuremath{CompTT_{\mathrm{T0}}} falls at the dips of the XIS pulse profile.
\item The powerlaw photon index ($\Gamma$) shows a strong dependence on the pulse profile. It is hardest at 
high energy PIN pulse profile peaks and softest at the off pulse phases.
Similar spectral hardening at the pulse peaks was found for this source previously by \citet{kreykenbohm1999}.
\item The folding energy (\ensuremath{E_{\mathrm{fold}}}) varies with the pulse phase with the values peaking at the ascending and descending phases of the PIN pulse profile,
and their minimum at the pulse peaks. The Cutoff-energy also varies with the pulse phase.
\item The electron temperature ($kT$) is highest at the ascending and descending phases of the PIN pulse profile
at phases $\sim$ 0.1, 0.5, 0.7 and 1. It is lower in the off pulse regions.
\item The optical depth ($\tau$) and $kT$ however shows strong anti-correlation with each other throughout the pulse phase. This may imply that the statistics of the 
data are not good enough to determine the variation of these two parameters independently as was already indicated
in the confidence contour maps.
\item Both the power law normalizations (\ensuremath{powerlaw_{\mathrm{norm}}}) and the CompTT normalizations (\ensuremath{CompTT_{\mathrm{norm}}}) shows a variability pattern with the values peaking at the peaks of the XIS pulse profile.
\end{enumerate} 

\section{Discussions \& Conclusions}

In the present work, we have performed a detailed pulse phase resolved analysis of a long \emph{Suzaku} observation of Vela X-1. Data from the same \emph{Suzaku} observation has been analyzed by \citet{doroshenko2011}. They
have probed the energy dependence of the pulse profile in four energy bands and have confirmed the presence
of the two CRSFs in the energy spectrum. We have however investigated the complex energy dependence of the pulse profiles
and probed the narrow dips of the same in much more detail by performing pulse phase resolved spectroscopy.
 Pulse phase resolved spectral analysis have been performed previously by \cite{kreykenbohm1999,kreykenbohm2002} using RXTE
 data. Apart from the variations in the continuum spectra with the pulse phase they also confirmed the existence of the two lines at $\sim$ 25 \& 
55 keV and reported 
their variation with the phase. In the present work, we have used a long observation of Vela X-1 with \emph{Suzaku}, and the high sensitivity
of \emph{Suzaku} over a broad energy band allows us to carry the same with much finer phase bins.
\cite{orlandini1998,labarbera2003} performed phase resolved analysis of Vela X-1
using \emph{BeppoSAX} data but could only confirm the existence of the absorption line at $\sim$ 55 keV. The results in all these
previous works are more or less in agreement with us. But we have been able to bring out these variations in more detail specially for the two 
CRSF features for which this is the most detailed result obtained so far. 

\subsection{Energy resolved pulse profiles}
Vela X-1 is a highly variable source and shows strong pulse to pulse variations. The low and high energy profiles 
have a very complex upto five peaked structure at lower energies which eventually merges to become a double peaked profile at higher energies.
The strong energy dependence of the pulse profiles have been probed previously by \citet{doroshenko2011} using data from the same \emph{Suzaku} observation
in four energy bands which show the structure of the complex pulse profiles. Their main emphasis was on comparing the pulse profiles between the on and ``off-states``. We have however investigated the 
complex energy dependence of the pulse profiles
and probed the narrow dips of the same in much more detail.
Presence of double peaked pulses at high energies have been found previously in some sources for example in XTE J1946+274 \citep{paul2001},
4U 1538-52 \citep{bildsten1997}, GX 301-2 \citep{koh1997} etc. They can either be interpreted due to a fan beam  pattern of the
emission geometry, or that due to the contribution from both the magnetic poles. The low energy complex pulse profiles are however more difficult 
to interpret since they
are affected due to scattering and absorption in the local environment of the neutron star. The dips in the low energy profiles can be explained
due to an additional absorption component in our line of sight, probably by the passage of a phase locked accretion stream or the accretion 
column itself. This aspect is discussed in more detail below.

\subsection{Phase resolved spectroscopy of the continuum spectrum}
We have performed pulse phase resolved spectroscopy to probe the complex energy dependence of the pulse profiles.
The dips in the pulse profiles at phases $\sim$ 0.3, 0.5-0.6, 0.7 and 0.9 can be explained due to an increase in the additional absorption column
density \ensuremath{N_{\mathrm{H2}}} with a corresponding change in the covering fraction at that phase. The changes in the value of \ensuremath{N_{\mathrm{H2}}} and covering fraction with
pulse phase can act as a tracer for the properties of the plasma in the accretion stream, which may be a clumpy structure having
different values of opacities and optical depths. The fitted parameters of the 'highecut' model shows
 an abrupt increase in the value of \ensuremath{N_{\mathrm{H2}}} at the dips at phases $\sim$ 0.3 and 0.9. The dip at phase $\sim$ 0.3 extends
upto somewhat higher energies, the value of \ensuremath{N_{\mathrm{H2}}} of the former being higher than that at phase $\sim$ 0.9. Both of
 the dips have similar strength in agreement with similar values of covering fraction in both the cases.
The main dip at phase 0.5-0.6 may be indicated by a slight increase in  \ensuremath{N_{\mathrm{H2}}} with a high value of the
covering fraction. The dip at phase $\sim$ 0.7 is different from 
the others since it increases in strength upto 5 keV and then decreases and disappears after 20 keV, being
relatively shallower than the other dips in the pulse profile. This dip is in agreement with a large
increase in \ensuremath{N_{\mathrm{H2}}} and a very small covering fraction, with the value of  \ensuremath{N_{\mathrm{H2}}} peaking at this 
particular phase. This would indicate the presence of a very dense clump of matter (a narrow and dense stream), with the high value of  \ensuremath{N_{\mathrm{H2}}} 
responsible for the dip extending upto higher energies. In general the covering fraction decreases at the dips
in the profile where \ensuremath{N_{\mathrm{H2}}} is larger, which may indicate a clumpy structure of the accretion stream. 
The 'highecut' model is thus able to explain the energy dependence
of the pulse profiles very well. The parameters obtained by fitting the spectra with
the 'CompTT' model on the other hand, does not show a similar pattern of variation of the partial covering model 
specially for the case of the 
covering fraction parameter. The dip at phase  $\sim$ 0.7 also does not show any significant increase in \ensuremath{N_{\mathrm{H2}}}, although the
optical depth ($\tau$) parameter of the 'CompTT' model increases at all phases corresponding to the dips. We would however like to point out that the 'CompTT' model may not be a good description of the low energy part of the pulsar spectra keeping in mind that the N$_{H2}$ and 
the covering fraction were found to have some dependence (mentioned earlier in section 4.2.2). In addition, it is a highly simplified model with a spherical shape of the Comptonizing cloud
whereas the emission region in accreting pulsars are likely to have a slab or cylindrical geometry.
The 'highecut' model inspite of being a phenomenological model and having a divergent property at low energies
when paired with a powerlaw, is able to explain the complex energy dependence of the pulse profile better.
 \newline Another interesting result found is the spectral hardening at the high energy pulse peaks. 
The power law index ($\Gamma$) reaches it's minima at these phases, more so for the phase corresponding to the second peak (phase $\sim$ 0.9). 
Similar regions of pulse hardening have been observed previously in other
 sources like Her X-1 \citep {pravdob, pravdoc}, 4U 0115+63 \citep {johnston1978, rose1979} \& GX 1+4 \citep {doty1981} \& 1A 1118-61 \citep{maitra2011}.
Assuming the spectral formation by the comptonization of photons by the thermal electrons, the Compton $Y$ parameter can be used 
as a measure for the degree of comptonization.  It is given by $y = \frac{4kT_{e}}{m_{e}c^{2}}max(\tau, \tau^{2}) $, where
$T_{e}$ is the electron temperature and $\tau$ the optical depth. 
\citep{rybici1986}. The higher is the degree of comptonization
the larger is the ''y'' parameter, and hence harder is the spectrum. "y" can be easily found since $kT_{e}$ and $\tau$
are the parameters of the 'CompTT' model. Calculating it shows that the "y" values also peak at more or less the same phase as 
 the hardest $\Gamma$, i.e at the pulse peaks
with the phase corresponding to the second peak having a higher value of "y". 
$y >>1$ at all the phases, which is indicative of saturated comptonization being prevalent in the the spectrum.
We also checked for the correspondence between ``$\Gamma$`` and "y" at all other phases. 
The parameters seem to be correlated with each other with the Spearman rank correlation coefficient equal to -0.89. Figure  \ref{fig10} shows the interdependence
between the two parameters.
 We would like to caution however that the 'CompTT'
model assumes spherical geometry for the plasma region which is a crude approximation for the accretion column which has a cylindrical or slab
geometry in most cases. Hence this may give rise to some systematic errors in the results.

\subsection{Phase resolved spectroscopy of the cyclotron parameters}
We also performed pulse phase resolved spectral analysis of the two CRSF features. Pulse phase resolved spectroscopy
of the cyclotron parameters have been performed for some sources previously, for example in Her X-1 \citep{soong1990,enoto2008,klochkov2008},
4U 1538-52 \citep{robba2001}, 4U0115+63 \citep{heindl2000}, Vela X-1 \citep{kreykenbohm1999,kreykenbohm2002,labarbera2003}, Cen X-3 \citep{suchy2008}, 
and more recently in GX 301-2 
\citep{suchy2012}, 1A 1118-61 \citep{suchy2011,maitra2011} and 4U 1626-67 \citep{iwakiri2012}. We have obtained detailed results on the variation of both
the CRSF's of Vela X-1 with the pulse phase. Both the 'highecut' and 'CompTT' models used by us to fit the spectra give comparable
results. This gives us more confidence on the obtained pattern of variation. The results are also very similar
to that obtained by \cite{kreykenbohm2002} with \emph{RXTE} but they did not have enough sensitivity to probe
these variations in detail. \newline
 There have also been attempts to model the CRSF features analytically by assuming
certain physics and geometry of the line forming region by \citet{araya1999,araya2000,schonherr2007} and more recently by 
\citet{nishimura2008,nishimura2011,mukherjee2011}. All these models predict significant variations in the depth, width and the centroid
energy of the CRSF features with the changing viewing angle at different pulse phases. They also predict possible deviation
or distortion from the simple dipole geometry of the magnetic field, and deviation of the line ratios from the classical value of two. 
Vela X-1 has a shallow and narrow fundamental, and a deep and wide harmonic (refer Table 1). This may happen due
to photon spawning by Raman Scattering at higher harmonics (refer \citep{schonherr2007}), or the superposition of a large
number of lines from different sites in the line forming region \citep{nishimura2008,nishimura2011} may also make
the fundamental CRSF appear shallower.
\cite{nishimura2011} further predicts a shallow and narrow
fundamental and a deeper and wider second harmonic would imply large viewing angle w.r.t the magnetic field. A fan like beam 
would be expected for Vela X-1 in this case. As is mentioned in section $4.2.1$ the ratio of the two line energies is $> 2$ at
many phases. According to \cite{nishimura2011}, This would favor a line forming region with a large polar cap or of a greater height. The fundamental would be formed at a higher site and second harmonic primarily around the bottom. As a result 
$\ensuremath{E2_{\mathrm{cycl}}}/\ensuremath{E1_{\mathrm{cycl}}} > 2$ is expected. \\
The electron temperature "$kT$" is high at the phases corresponding to the wider CRSFs, specially at
the ascending edges of the high energy PIN pulse peaks. For a Maxwell-Boltzmann distribution of electrons the observed
CRSF FWHM is given as :
\citep{meszaros1992} $W_{c} \propto E_{C}\sqrt{kT_{e}}|\cos \theta|$, \newline
where $kT_{e}$ is the electron temperature, $E_{C}$ the CRSF energy, and $\cos \theta$ the viewing angle w.r.t the magnetic field.
This implies a higher $kT_{e}$ would result into a higher $W_{c}$ as is obtained. 
The fundamental line is deepest at the ascending edge of the pulse phase. The deepest 
second harmonic is also in the  descending edge of the main pulse and very shallow outside the pulses. \cite{nishimura2011} predicts these signatures would imply looking at a very large viewing angle w.r.t the magnetic field at those phases and hence a fan beam
like emission pattern would be expected.
All these are however assuming the contribution from one of the magnetic poles of the neutron star, and results
may be complicated further with the contribution from the other pole.
 
\acknowledgments

This research has made use of data obtained through the High Energy Astrophysics Science Archive Research Center On line Service,
 provided by NASA/Goddard Space Flight Center.

\clearpage

\clearpage
\clearpage
\begin{figure}
\begin{center}
\includegraphics[height=10 cm]{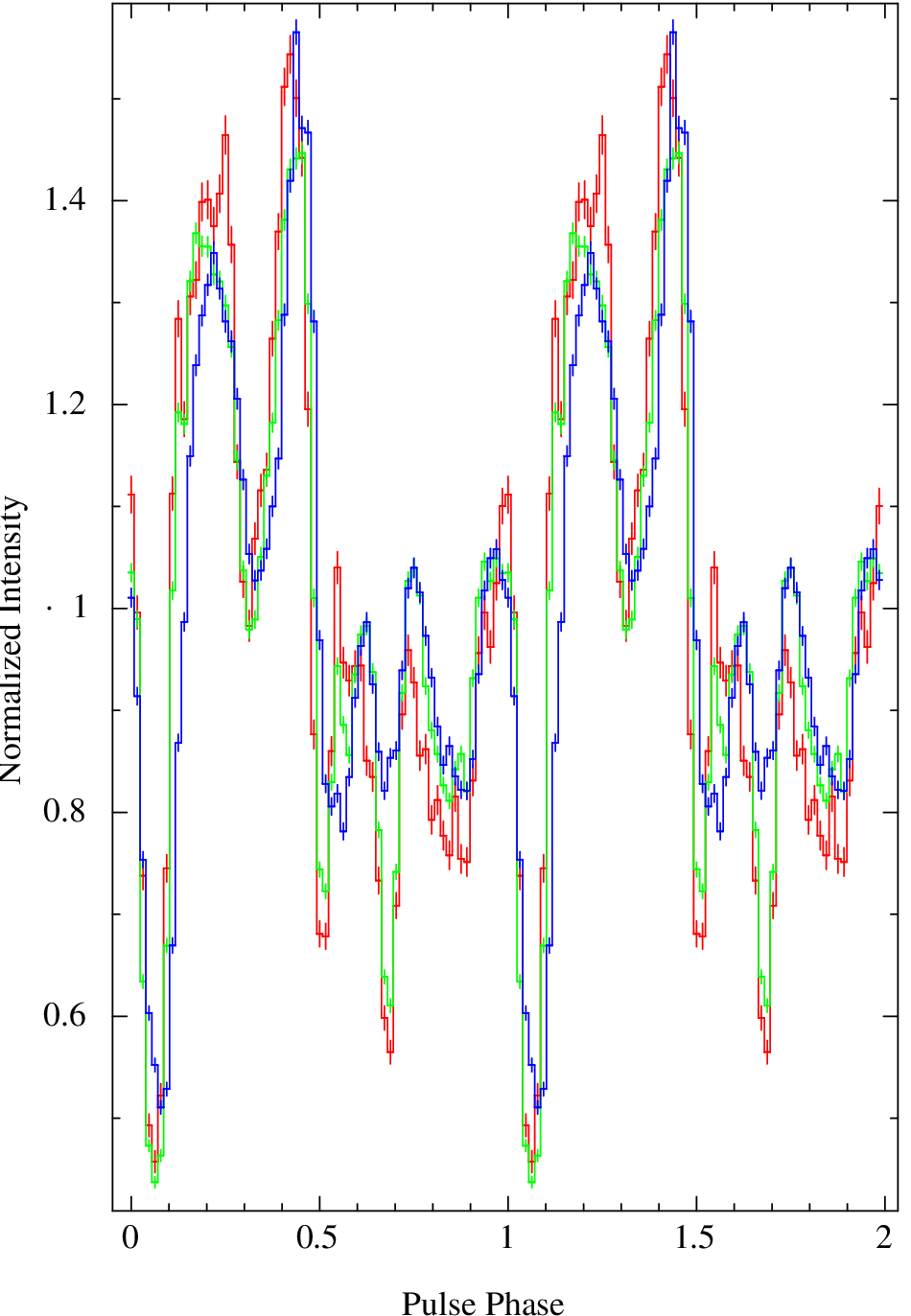}
\includegraphics[height=10 cm]{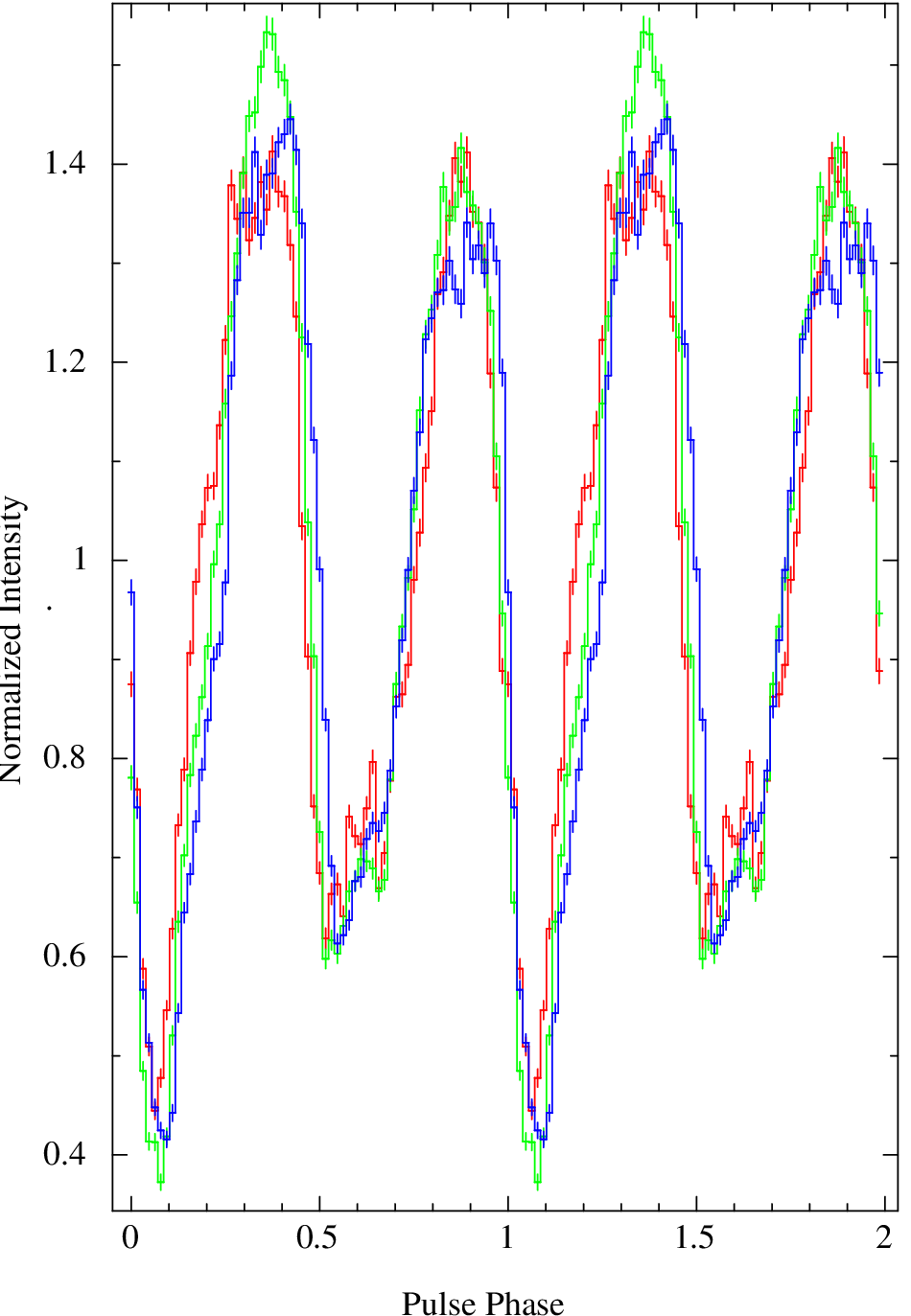}
\caption{Intensity dependent pulse profiles of Vela X-1. The left panel shows the pulse profiles
(0.3-12 keV) using XIS data \& the right panel shows the pulse profile (10-70 keV) using PIN data.
The pulse profiles in red shows the profile with XIS count rate $\le$ 20 c $s^{-1}$, the one in green with 20-50 c $s^{-1}$, and the one in blue $\ge$ 50 c $s^{-1}$}
\end{center}
\label{fig3}
\end{figure}

\begin{figure}
\includegraphics[height=8 cm]{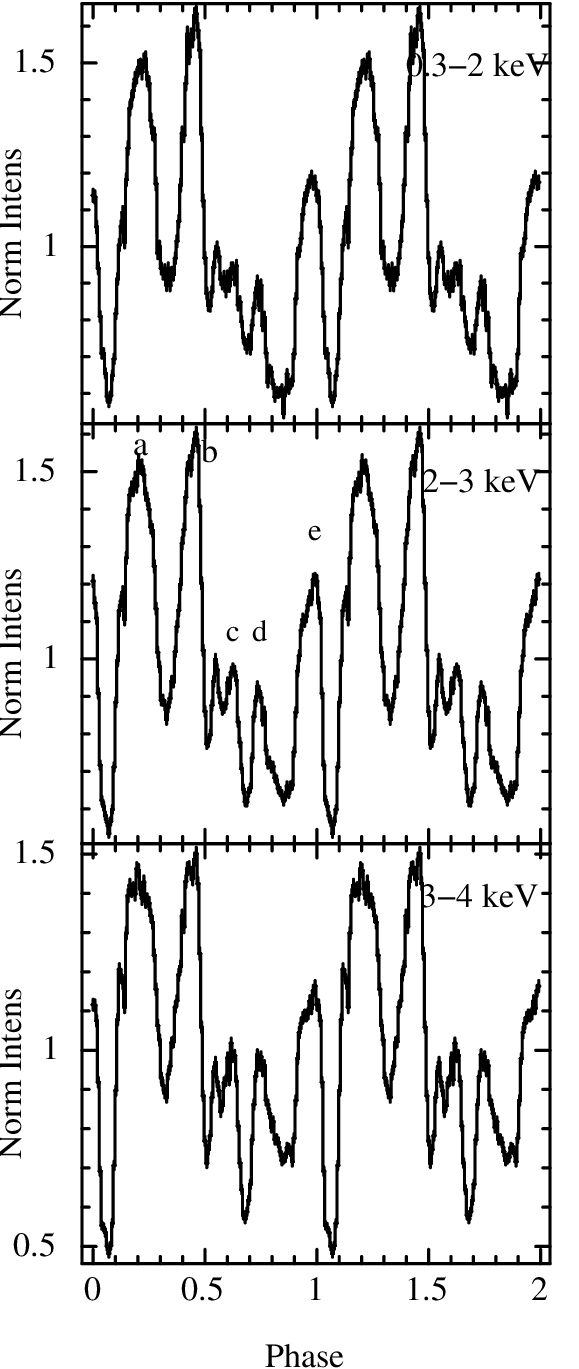}
\includegraphics[height=8 cm]{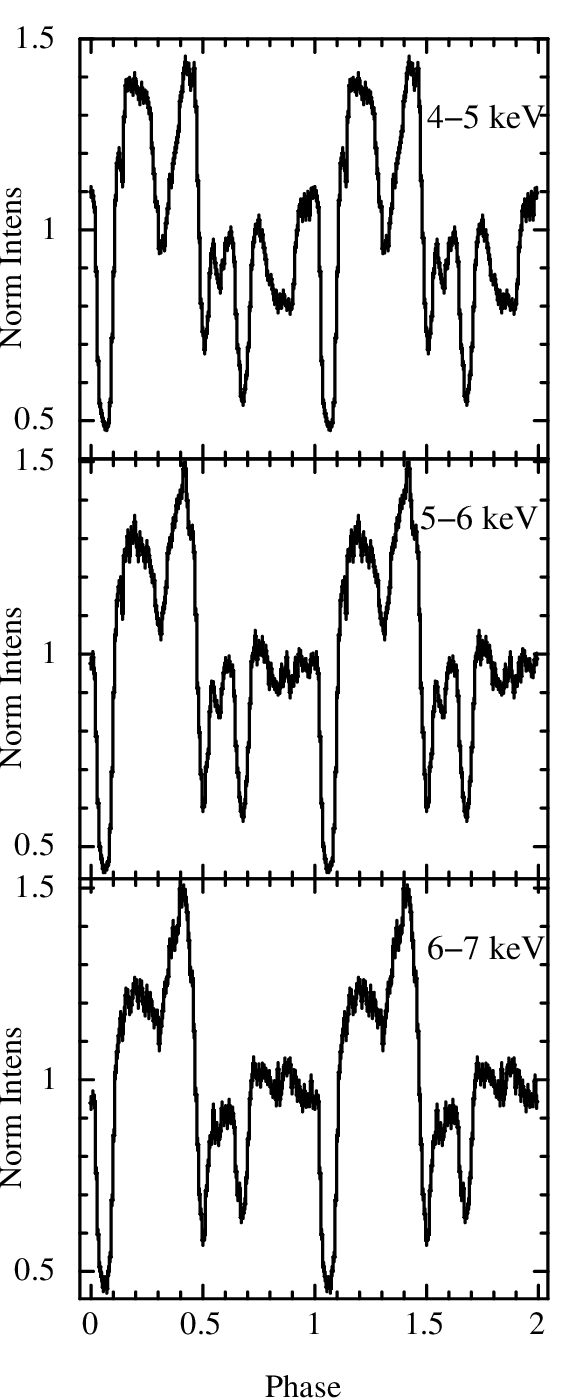}
\includegraphics[height=8 cm]{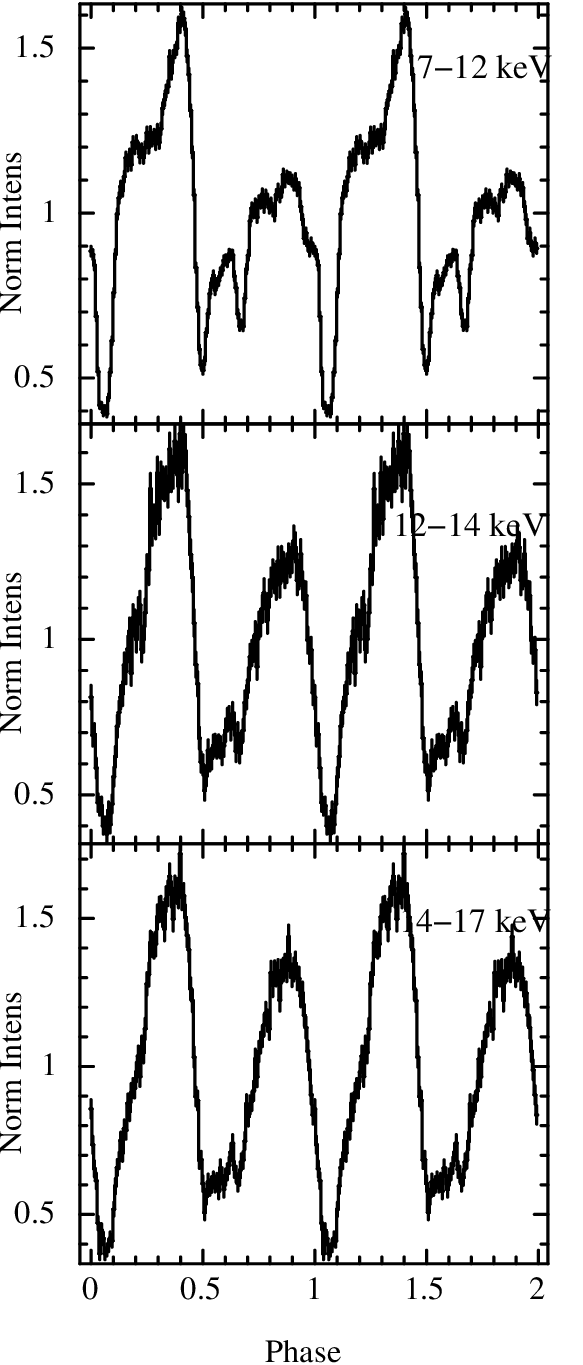}
\includegraphics[height=8 cm]{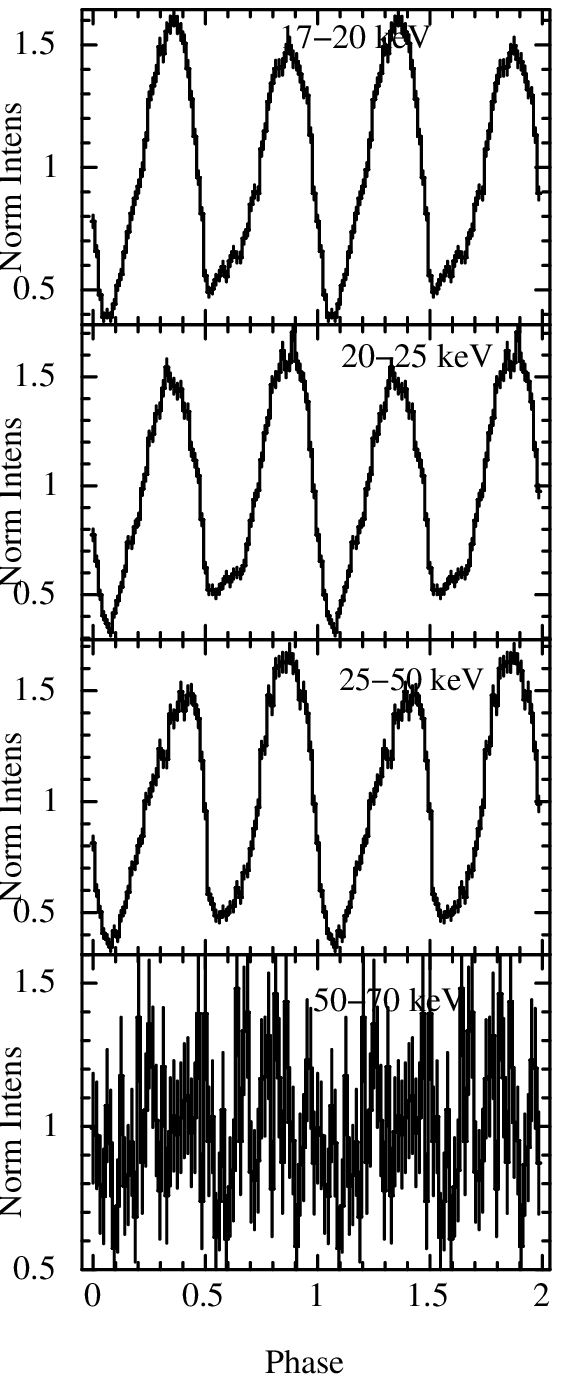}
\caption{Energy dependent pulse profiles of Vela X-1 using XIS \& PIN data. 
The energy range for the pulse profiles are specified inside the panels.}
\label{fig4}
\end{figure}
\begin{figure}
\includegraphics[scale=0.35]{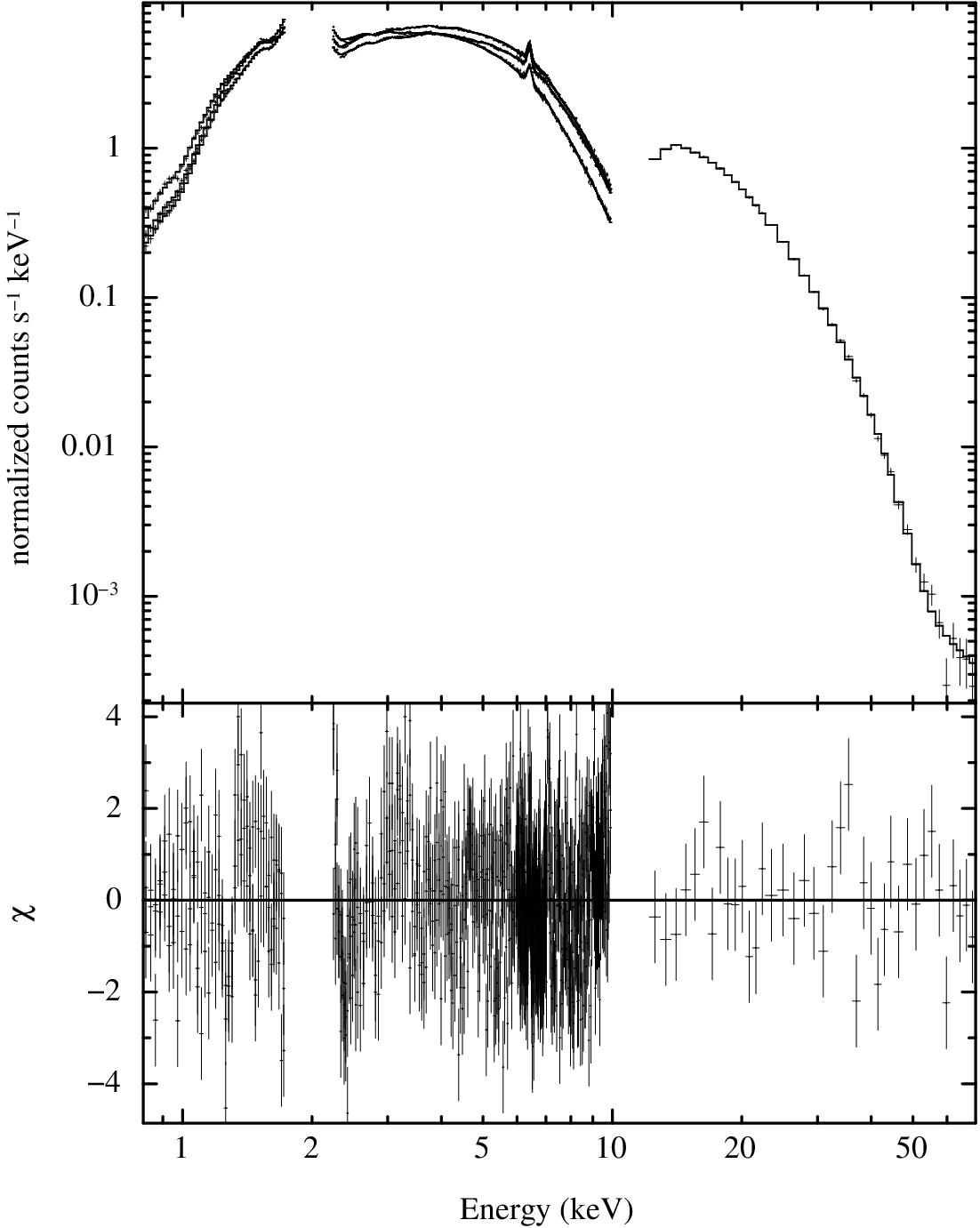}
\includegraphics[scale=0.35]{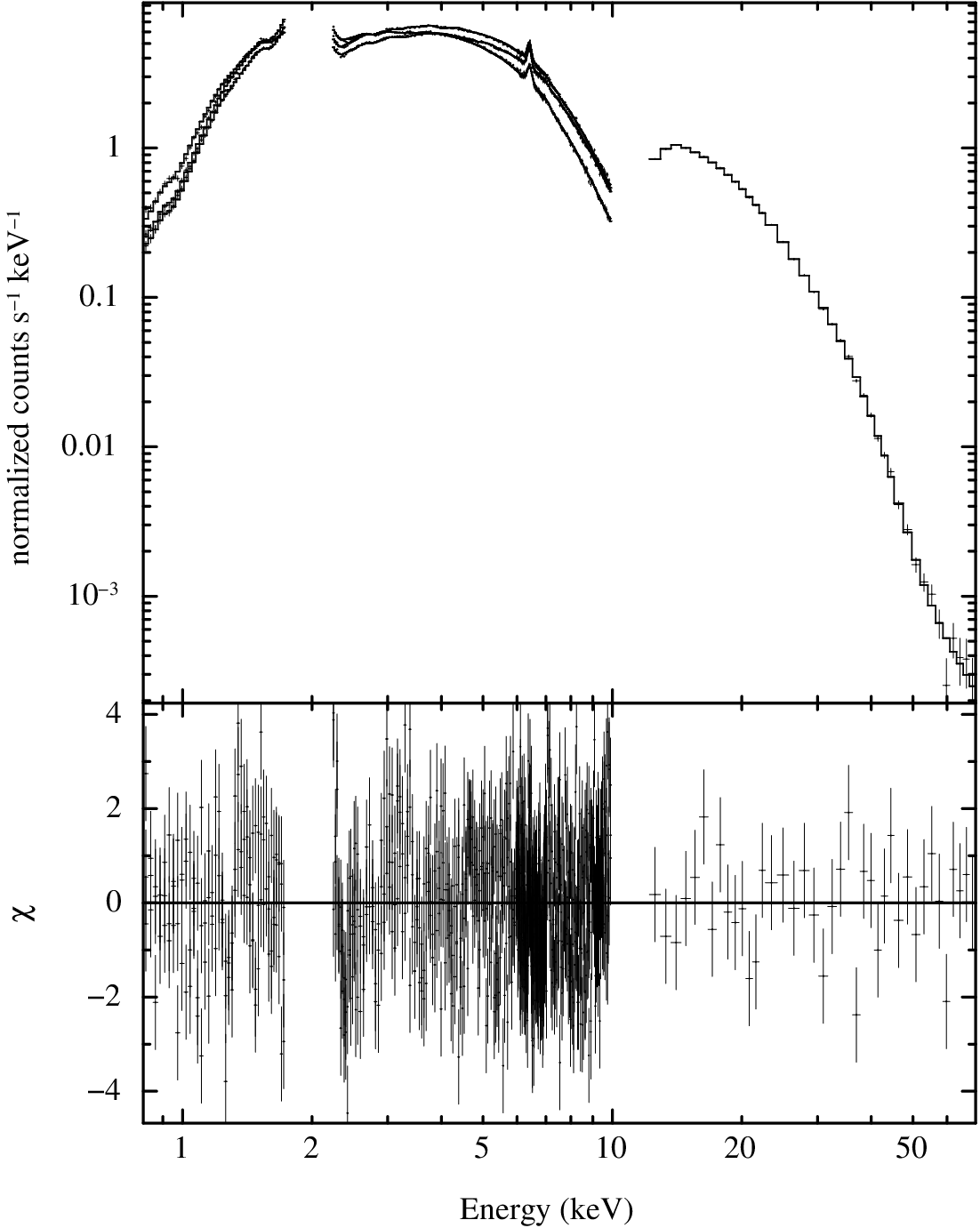}
\caption{The pulse-phase averaged spectrum of Vela X-1 using high-energy cutoff model (left panel) and 'CompTT' model (right panel).
The upper panel shows the best-fit spectra. The residuals are given in the bottom panels.}
\label{fig5}
\end{figure}

\clearpage
\begin{figure}
 \scriptsize 
\begin{center}$
\begin{array}{ccc}
\includegraphics[height=5 cm,angle=-90]{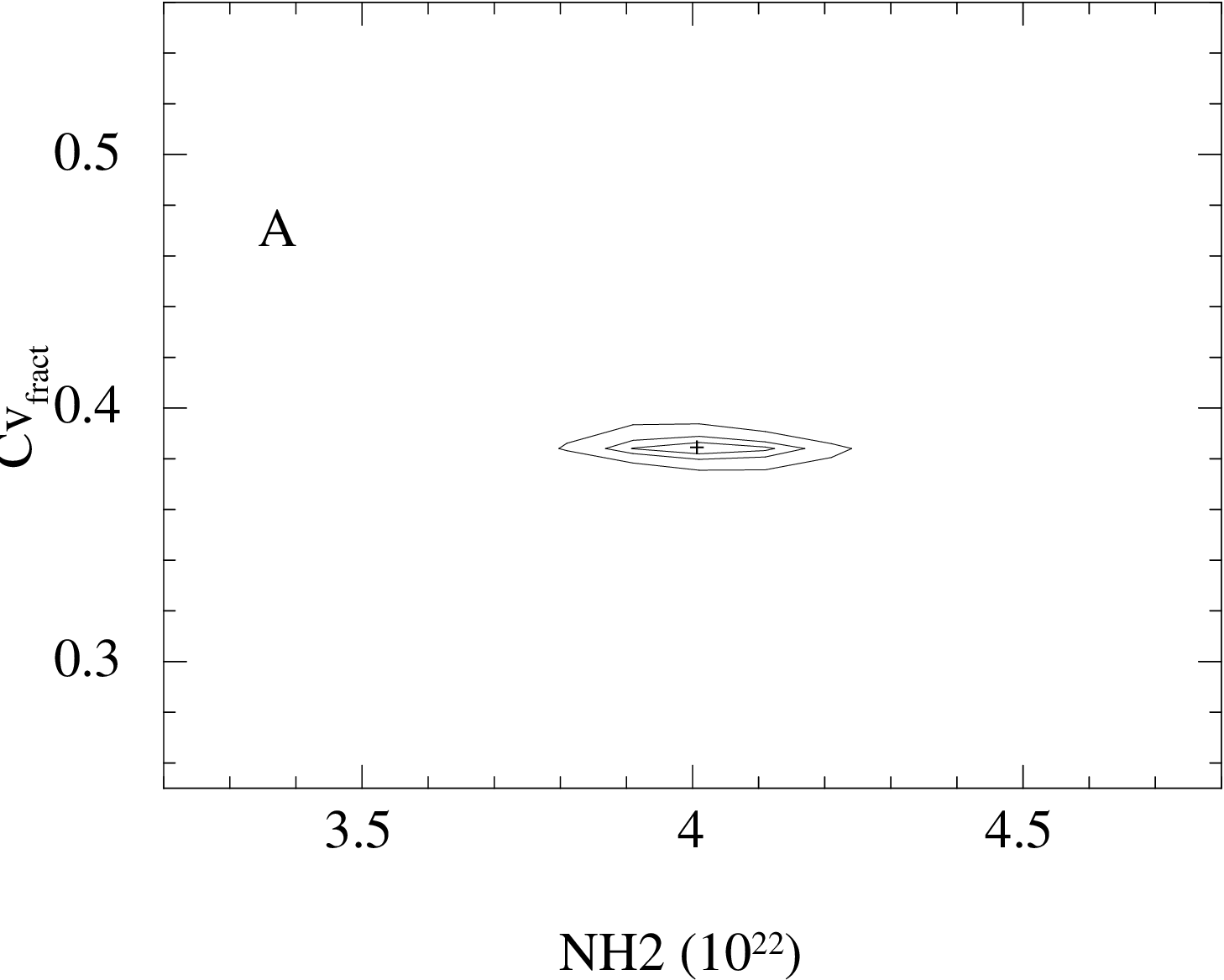} &
\includegraphics[height=5 cm,angle=-90]{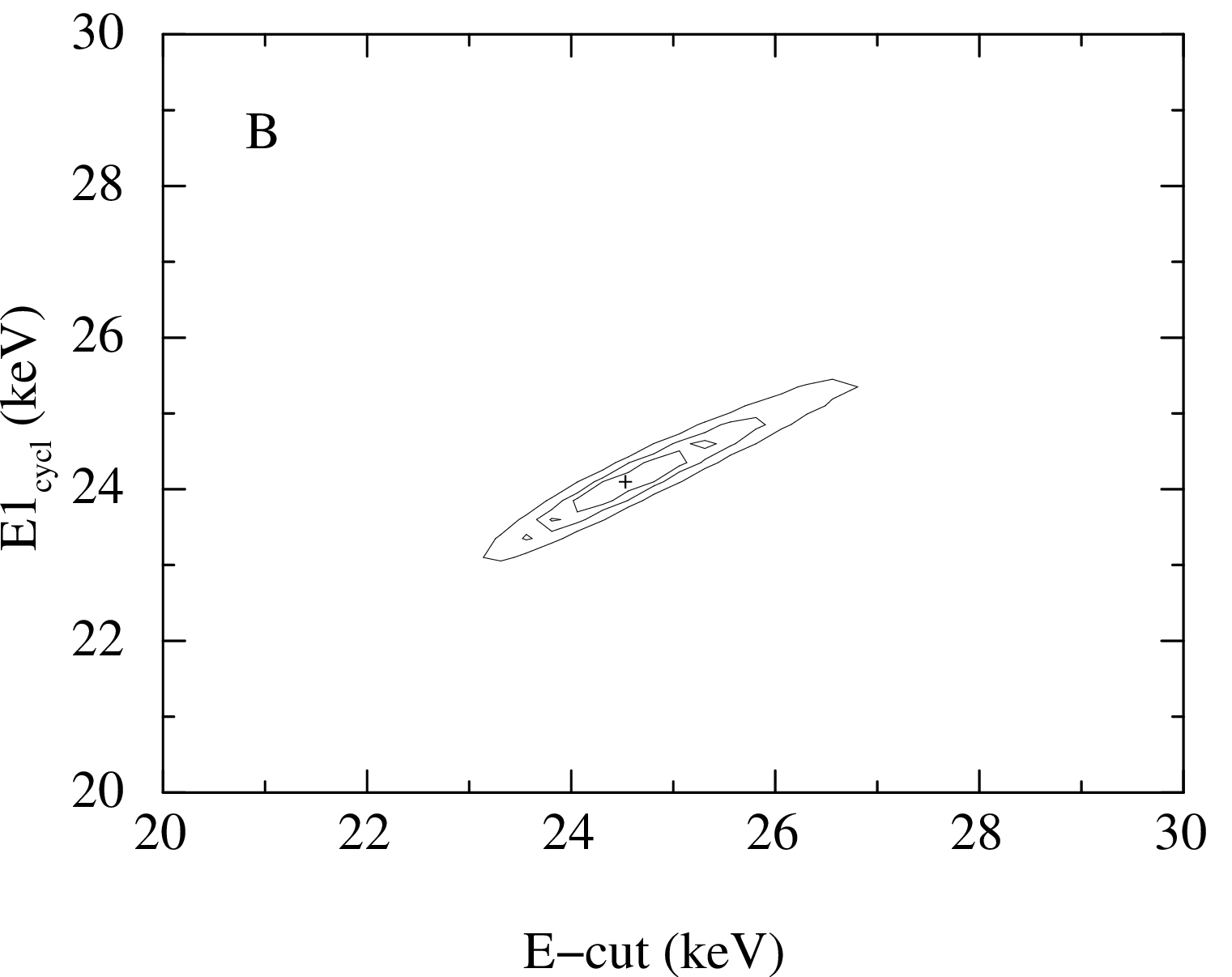} &
\includegraphics[height=5 cm,angle=-90]{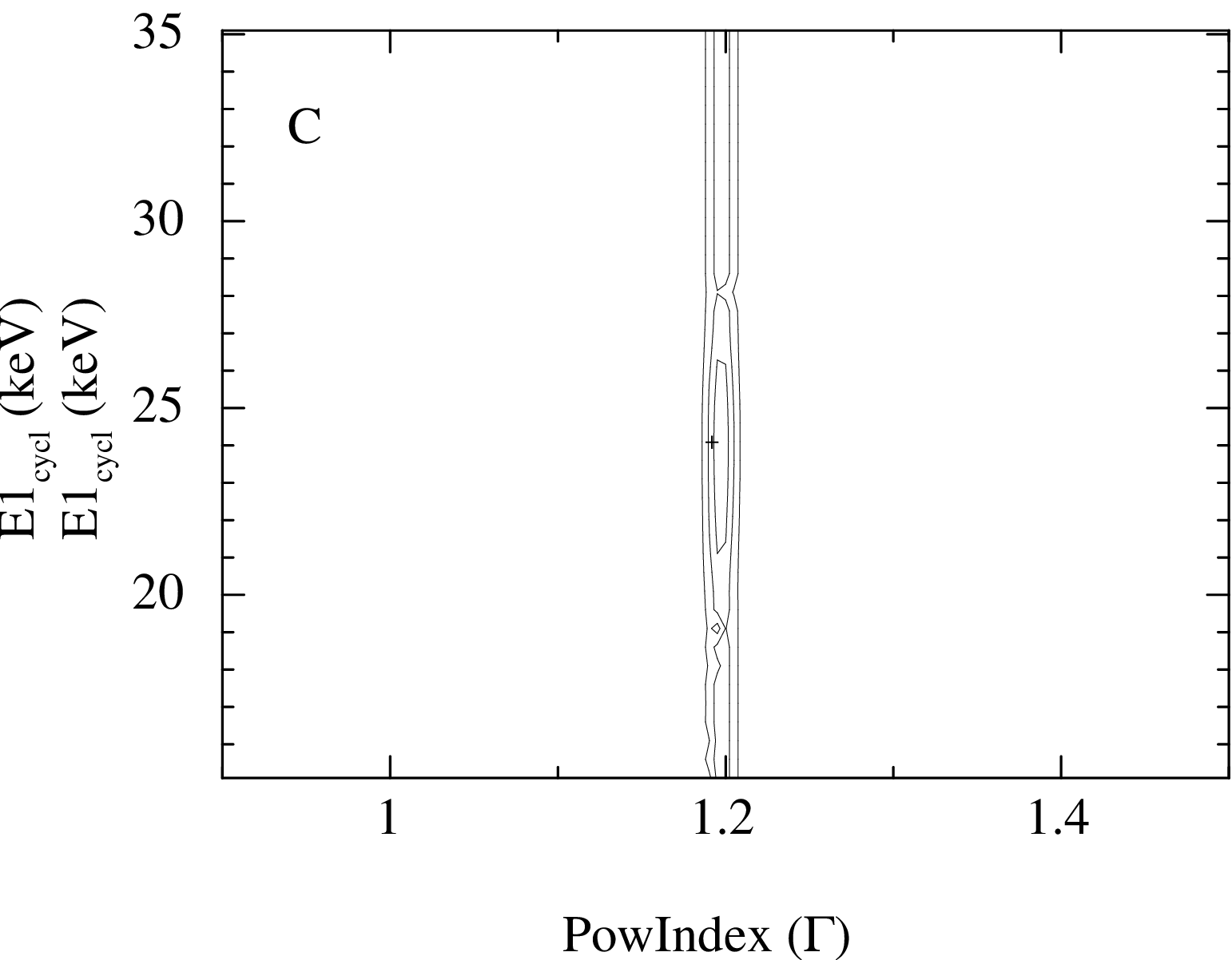} \\
\includegraphics[height=5 cm,angle=-90]{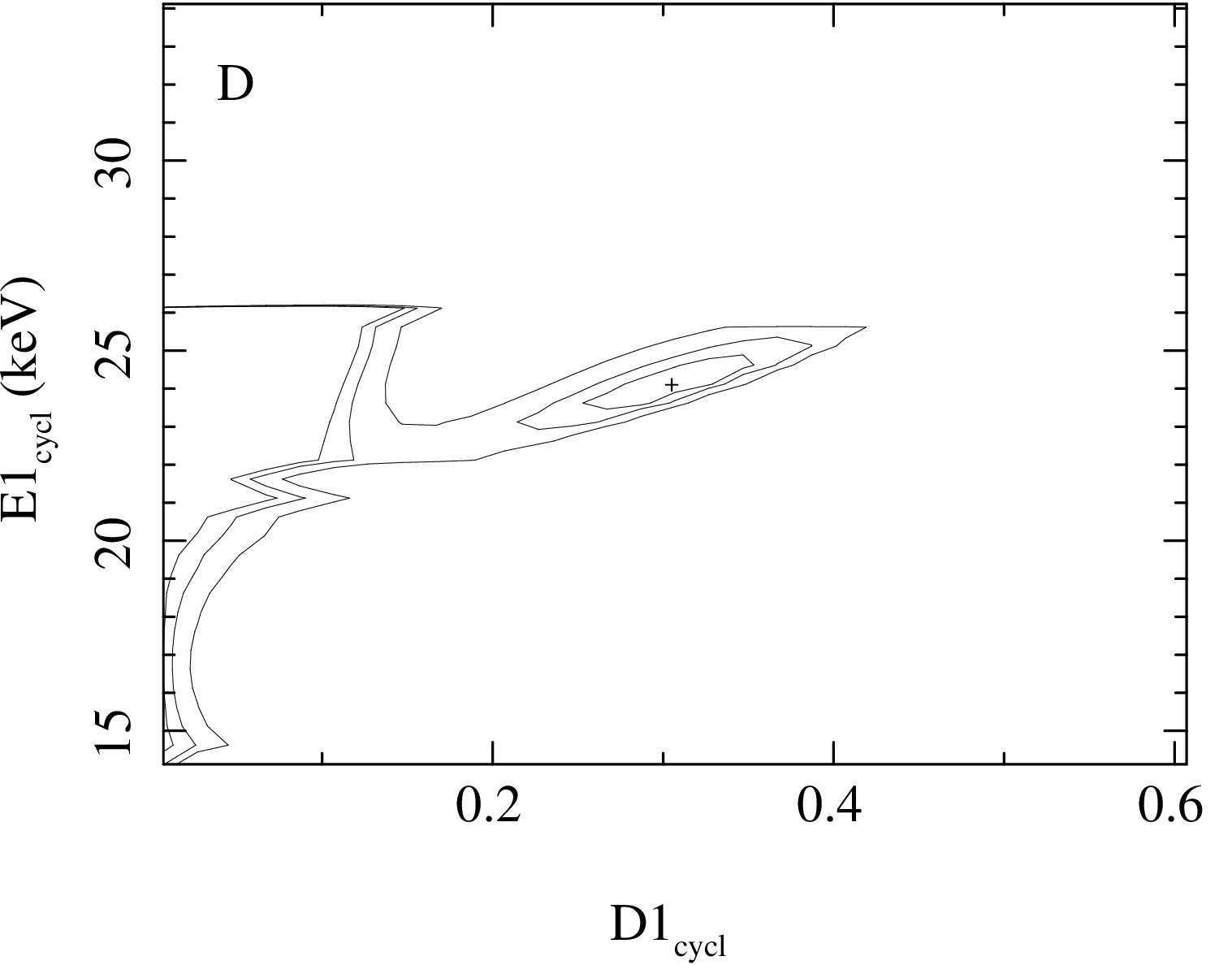} &
\includegraphics[height=5 cm,angle=-90]{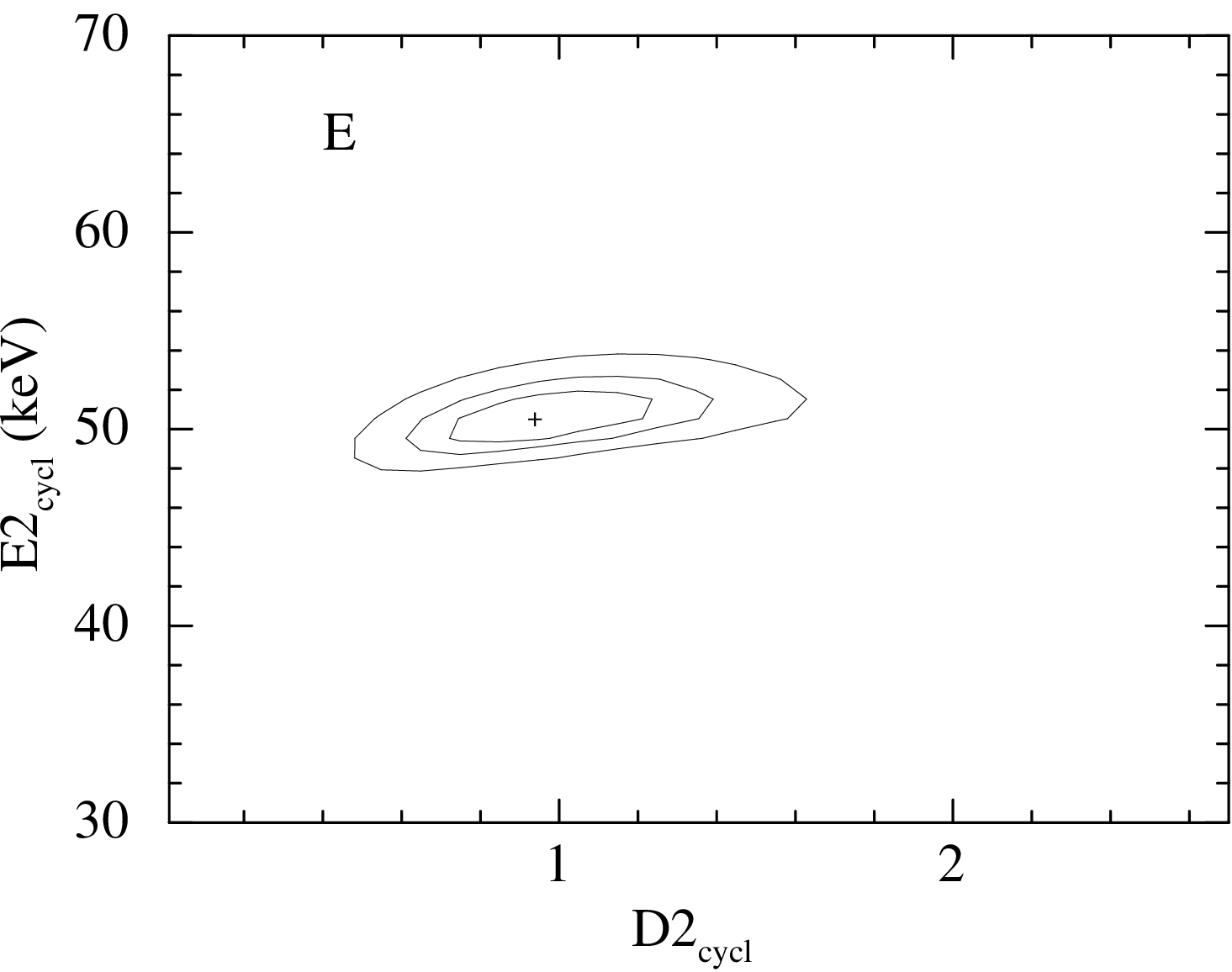} \\ 
 &  &  \\
\hline
\newline
\includegraphics[height=5 cm,angle=-90]{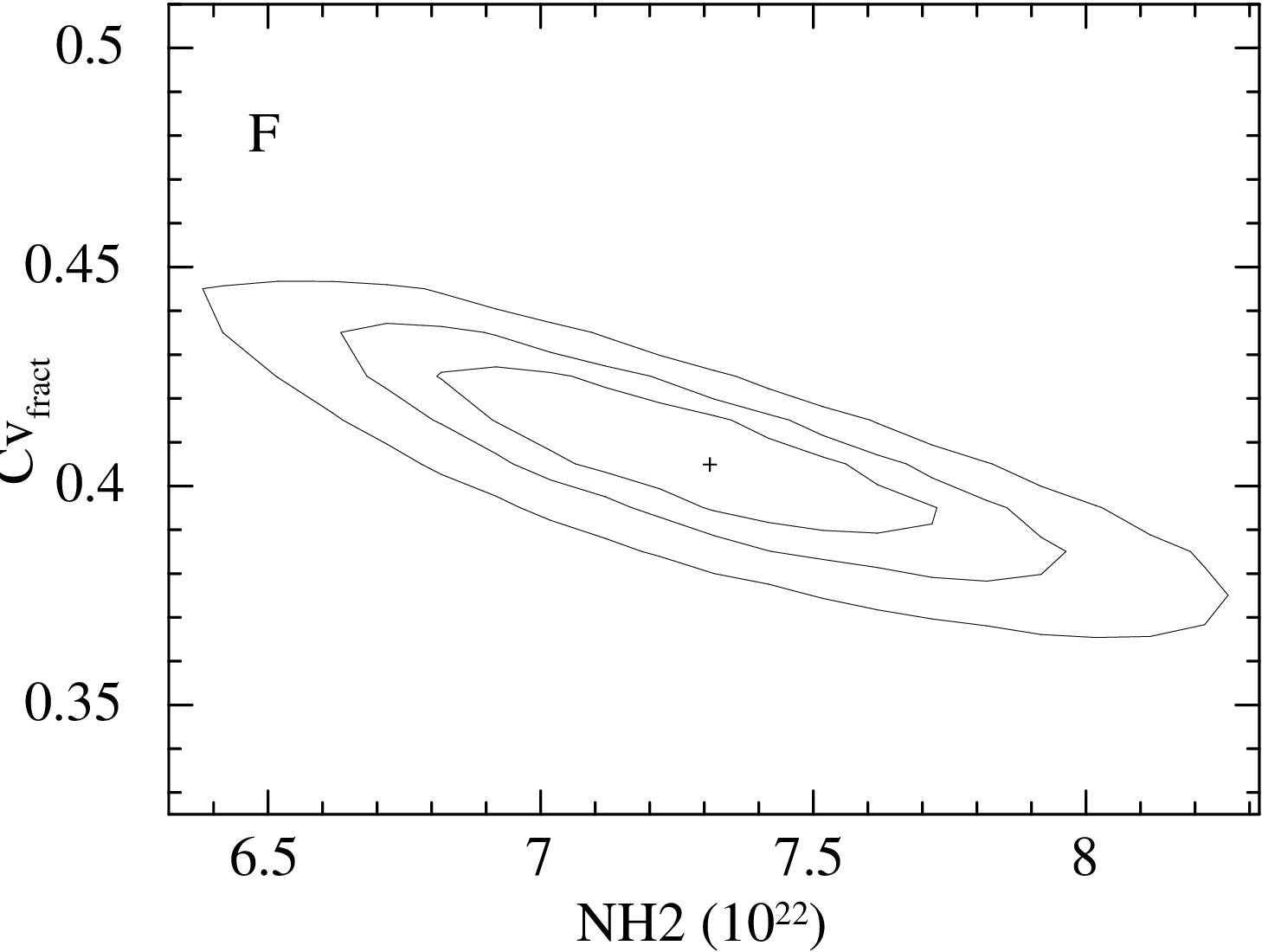} &
\includegraphics[height=5 cm,angle=-90]{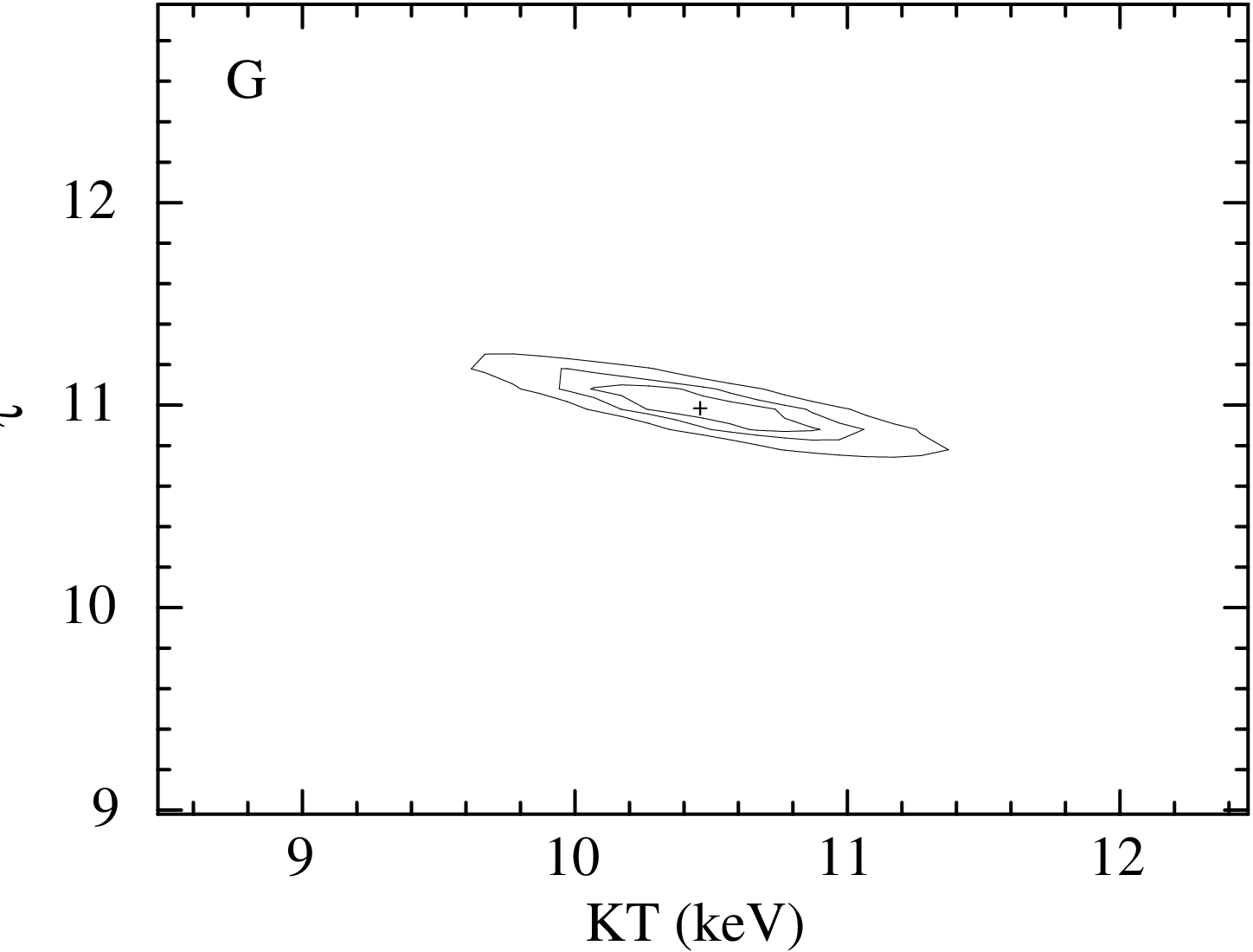} &
\includegraphics[height=5 cm,angle=-90]{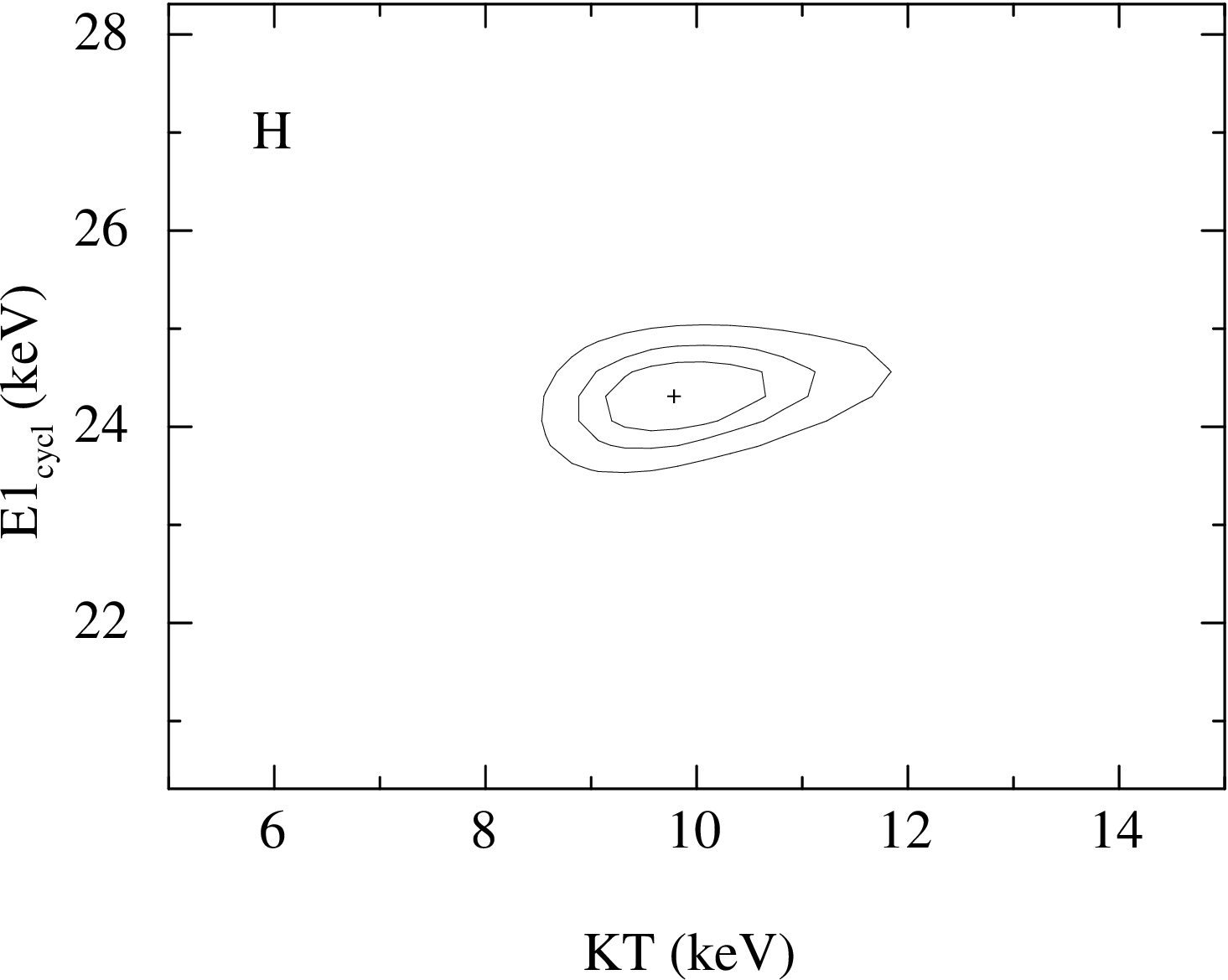} \\
\includegraphics[height=5 cm,angle=-90]{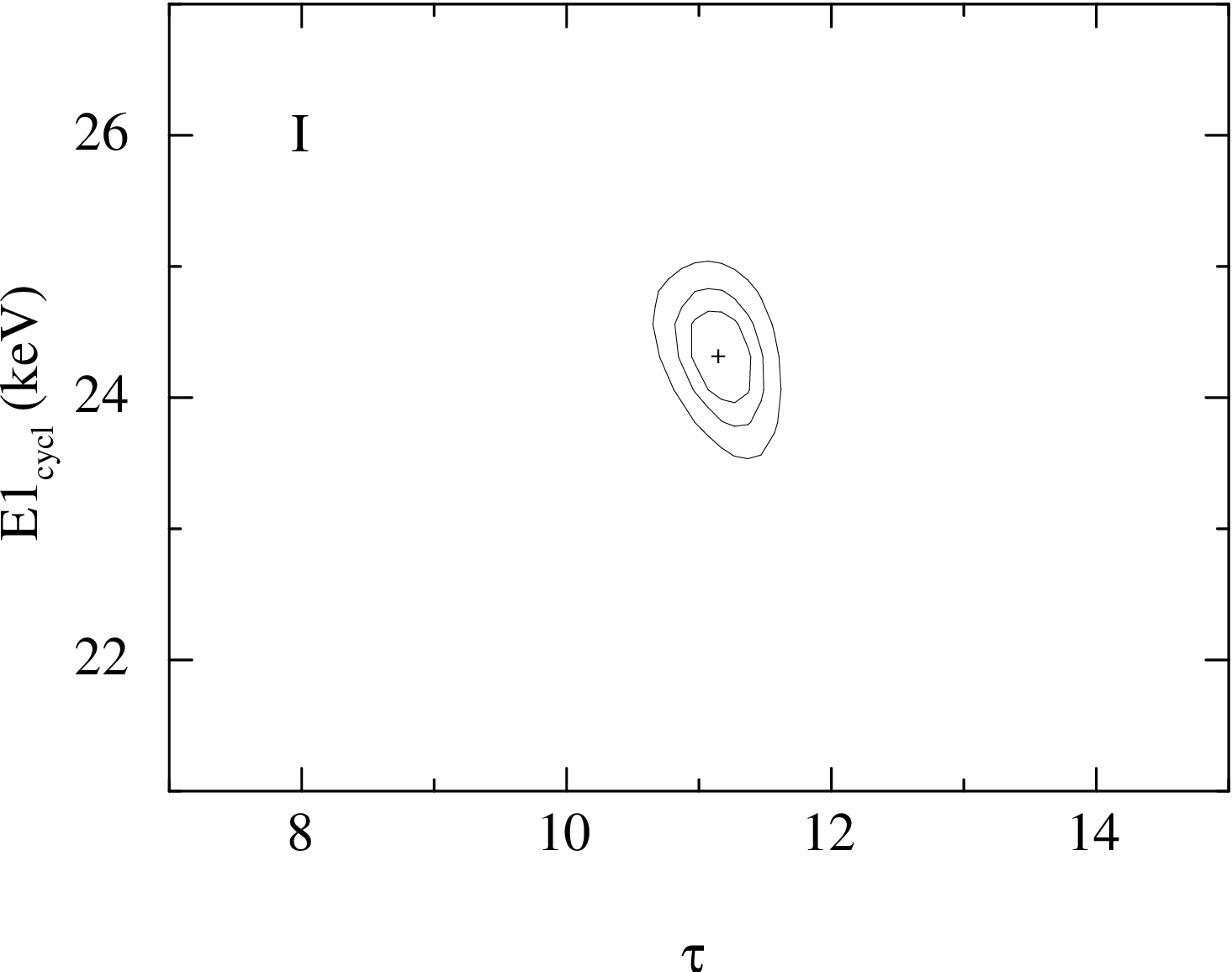} &
\includegraphics[height=5 cm,angle=-90]{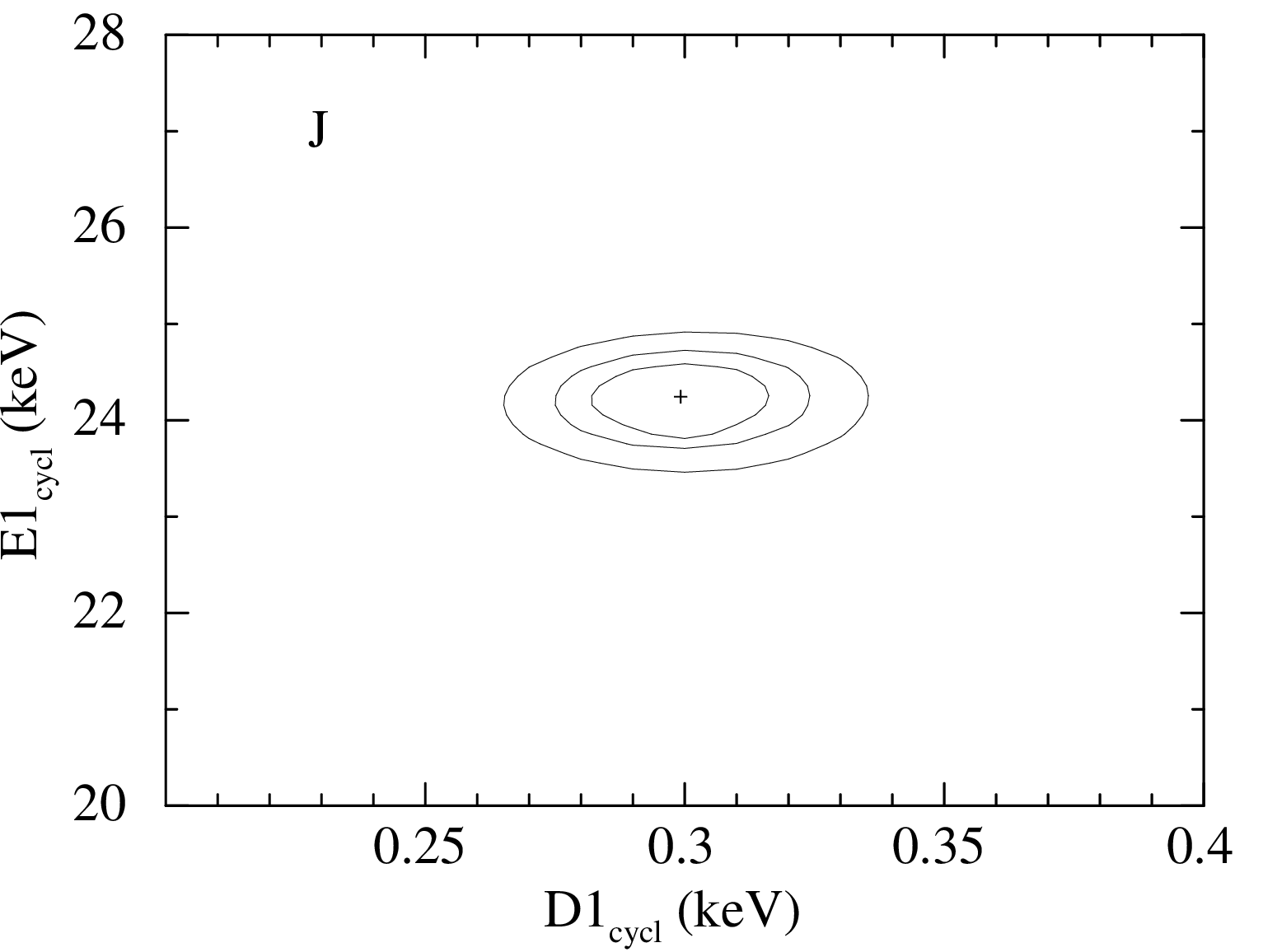} &
\includegraphics[height=5 cm,angle=-90]{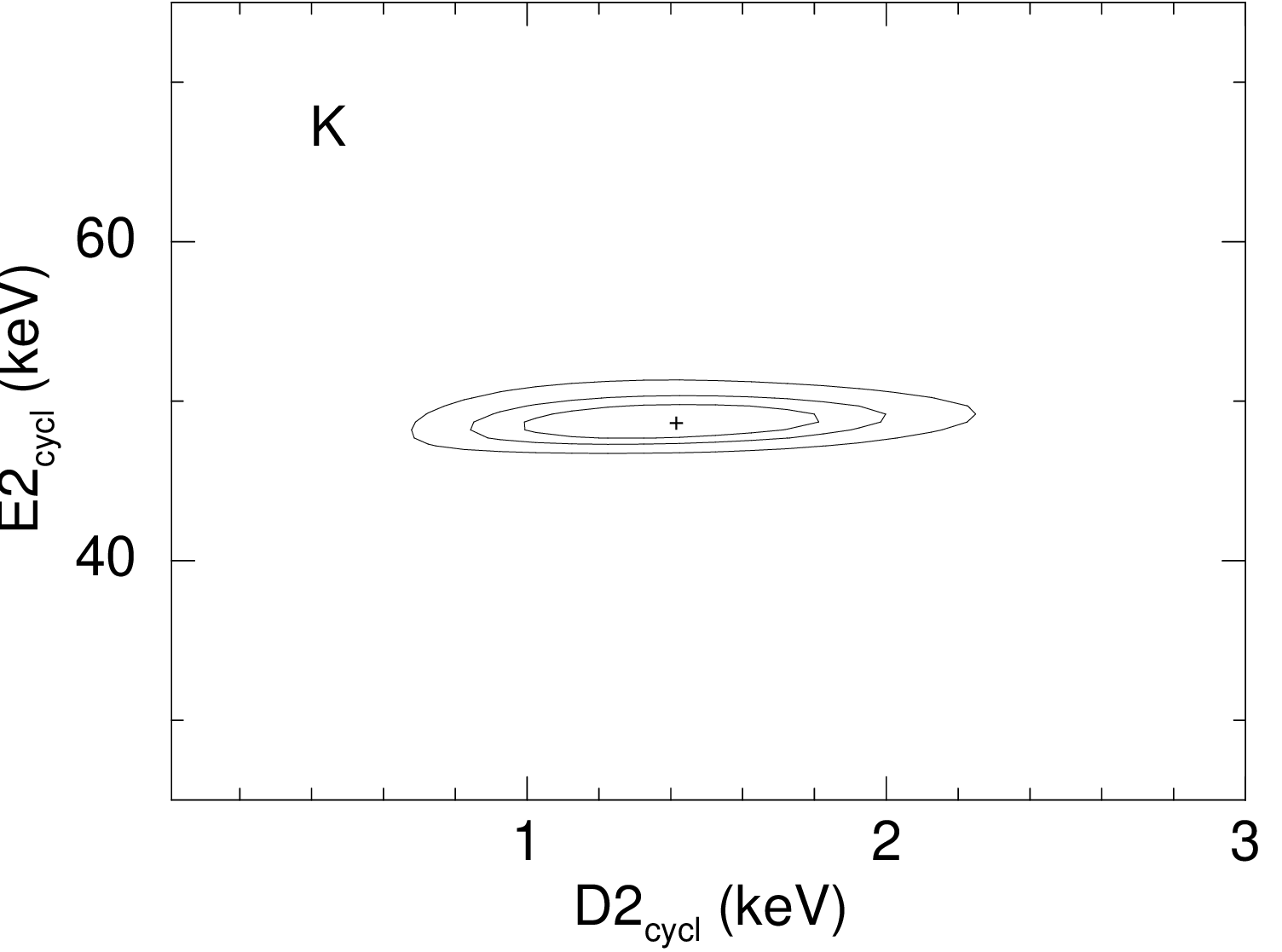} 
\end{array}$
\end{center}
\caption{\scriptsize
$\chi^{2}$ confidence contours between several pairs of continuum and line parameters
for the two models, obtained from the phase averaged spectra. The innermost to outermost contours
represent respectively 68$\%$, 90$\%$ and 99$\%$ confidence levels. The upper two panels
separated with a horizontal line show the contour plots for the 'highecut' model and the lower
panels for the 'CompTT' model.
The plots show the confidence
contours between A)\ensuremath{N_{\mathrm{H2}}} and \ensuremath{Cv_{\mathrm{fract}}} B) E-cut energy and \ensuremath{E1_{\mathrm{cycl}}} C) PowIndex and \ensuremath{E1_{\mathrm{cycl}}}
D) \ensuremath{E1_{\mathrm{cycl}}} and \ensuremath{D1_{\mathrm{cycl}}} \& E) \ensuremath{E2_{\mathrm{cycl}}} and \ensuremath{D2_{\mathrm{cycl}}}, for the 'highecut' model, and
F) \ensuremath{N_{\mathrm{H2}}} and \ensuremath{Cv_{\mathrm{fract}}} G) $\tau$ and $kT$ H) \ensuremath{E1_{\mathrm{cycl}}} and $kT$ I)  \ensuremath{E1_{\mathrm{cycl}}} and  $\tau$ J) \ensuremath{E1_{\mathrm{cycl}}} and \ensuremath{D1_{\mathrm{cycl}}} \&
K)  \ensuremath{E2_{\mathrm{cycl}}} and \ensuremath{D2_{\mathrm{cycl}}} for the 'CompTT' model.}
\label{fig11}
\end{figure}

\begin{figure}
\centering
\includegraphics[height=12cm, width=9cm]{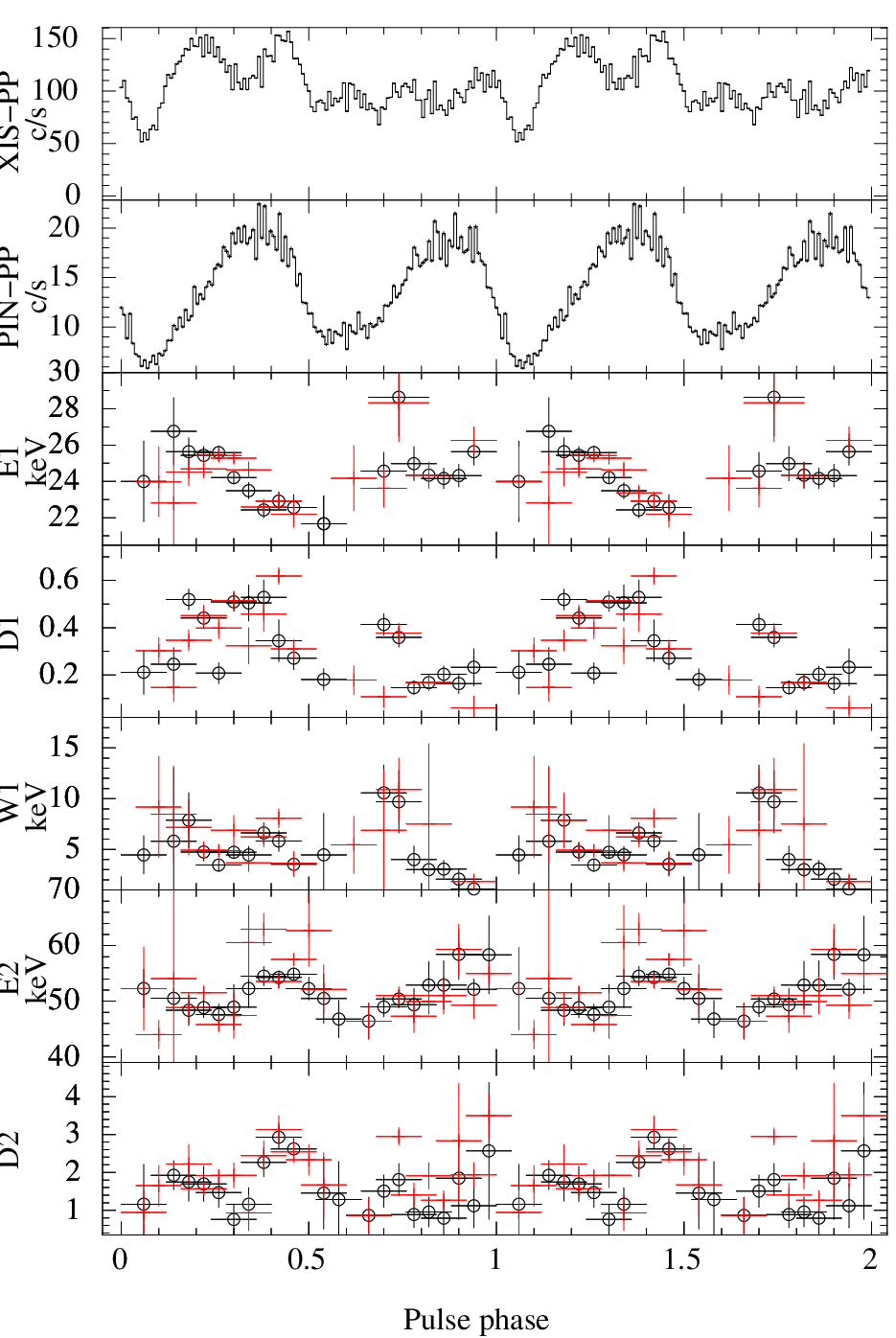}
\caption{Variation of the cyclotron line parameters as is obtained with the two models. The black points 
with 'circle' denotes the parameters as obtained with the highecut model The red points denote the
points as obtained with the CompTT model. Only 8 of the 25 bins are indepenent.
Although the parameters for the two models match reasonably well,
the fundamental line energy for the Highecut model cannot be constrained well and depends on the continuum parameters, specially the cutoff energy as shown in the contour plots in Figure \ref{fig11}.}
\label{fig6}
\end{figure}

\begin{figure}
\includegraphics[scale=0.6,angle=-90]{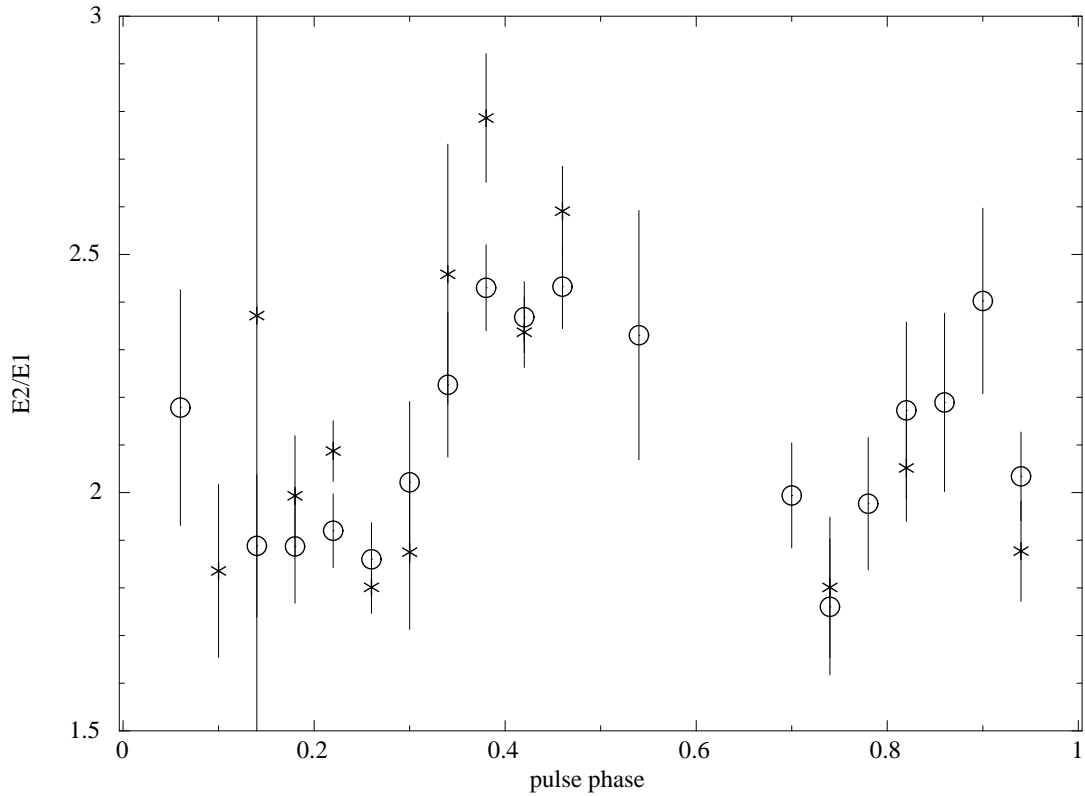}
\caption{Variation of the ratio of the two cyclotron lines E2/E1 as is obtained by fitting the two models. The variation of the parameter as obtained from the high energy cutoff model is denoted by the symbol 'circle'. Variation of the parameter as obtained from the CompTT model is denoted by the symbol 'star'.}
\label{fig9}
\end{figure}

\begin{figure}
\includegraphics[height=12 cm, width=8 cm
]{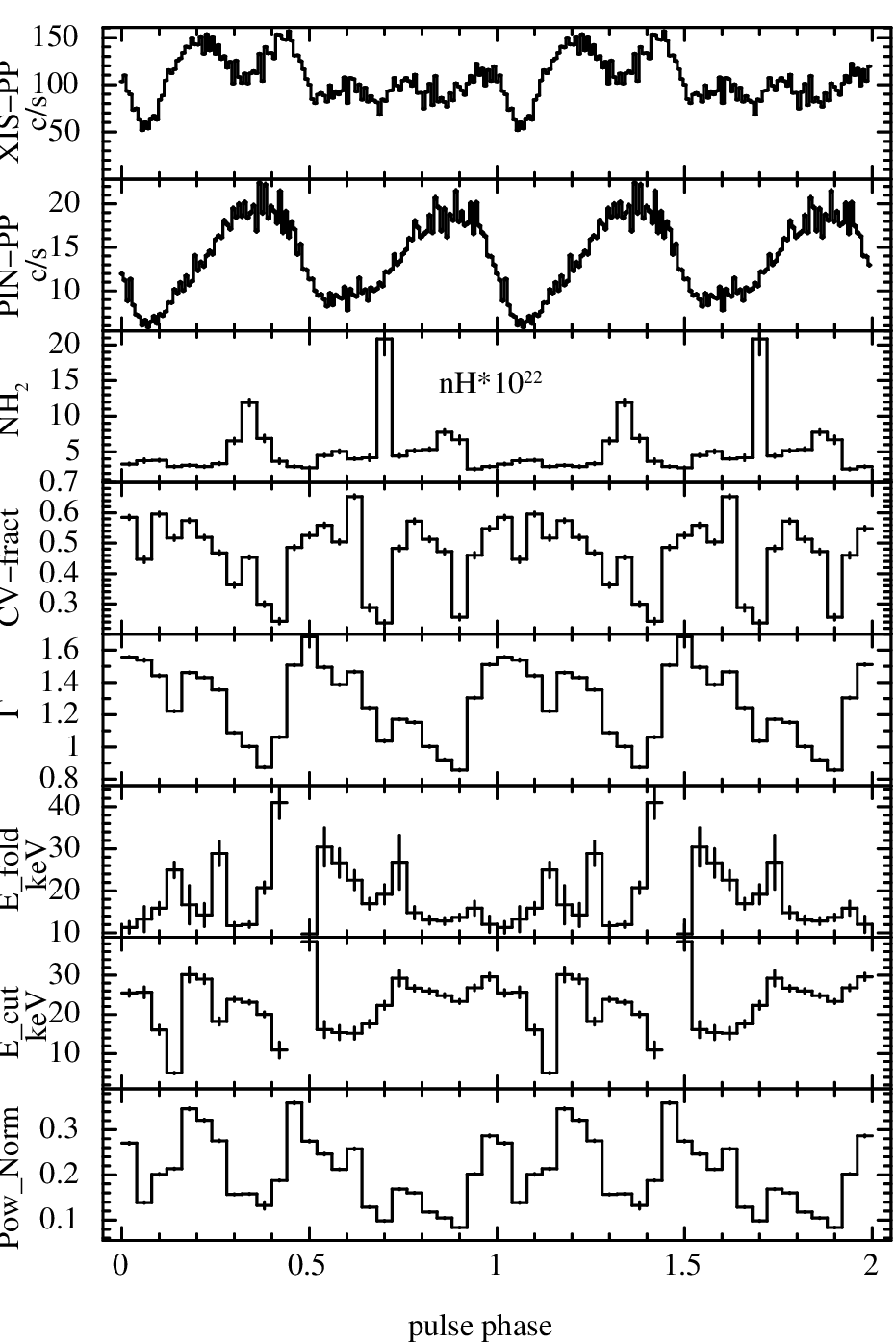}
\includegraphics[height=12 cm, width=8 cm]{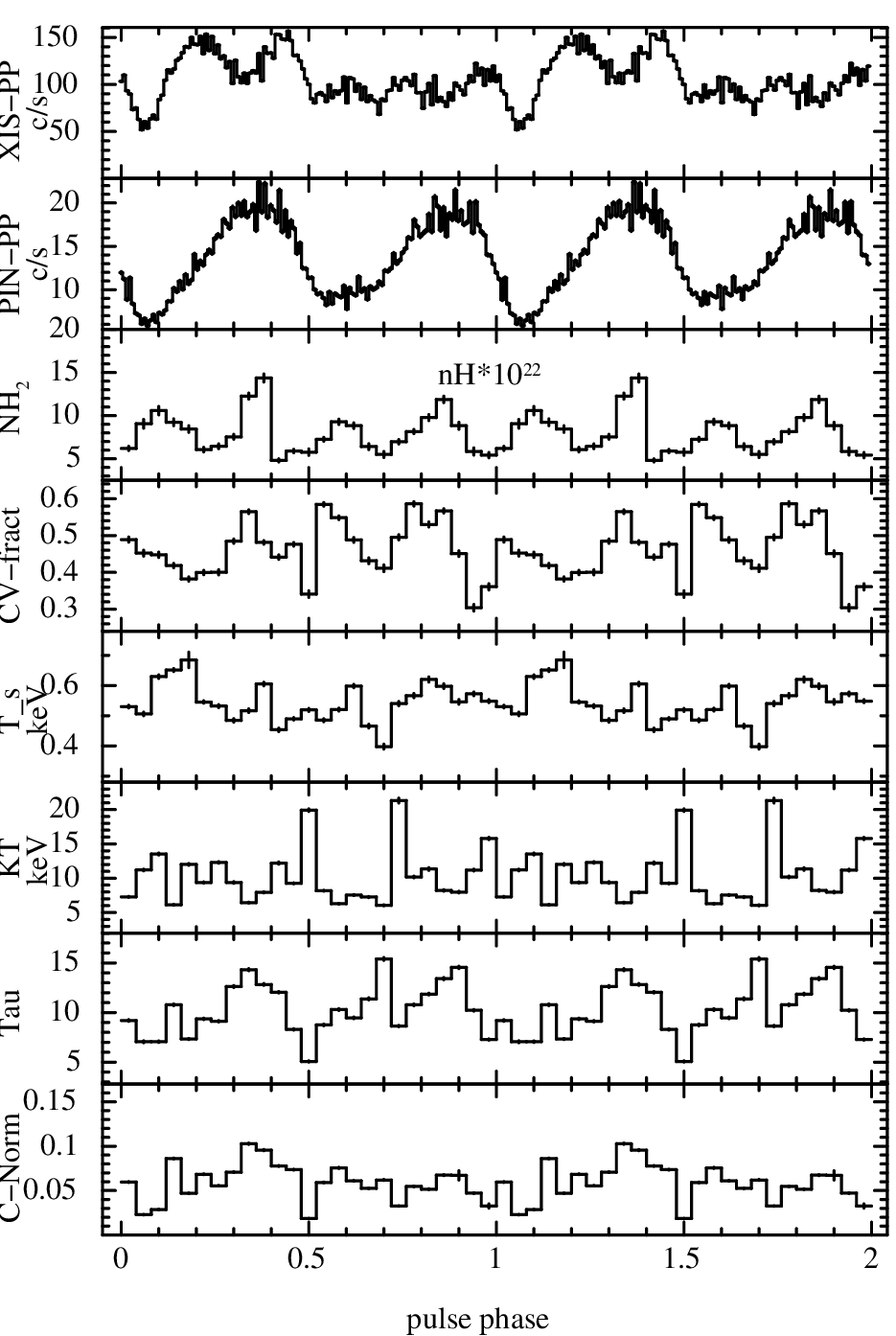}
\caption{Variation of the spectral parameters with phase along with the pulse profile (0.3-12 keV for XIS and 10-70 keV for PIN). 
The left panel shows the variation  
using the model 'highecut' and the right panel shows the same using the model 'CompTT'.
For the 'highecut' model, the real uncertainty in the value the cutoff energy is bigger than shown in the plot considering it's dependence
on the fundamental line energy as seen in the contour plots (Figure \ref{fig11})}.
\label{fig7}
\end{figure}
\clearpage
\begin{figure}
\includegraphics[scale=0.6,angle=-90]{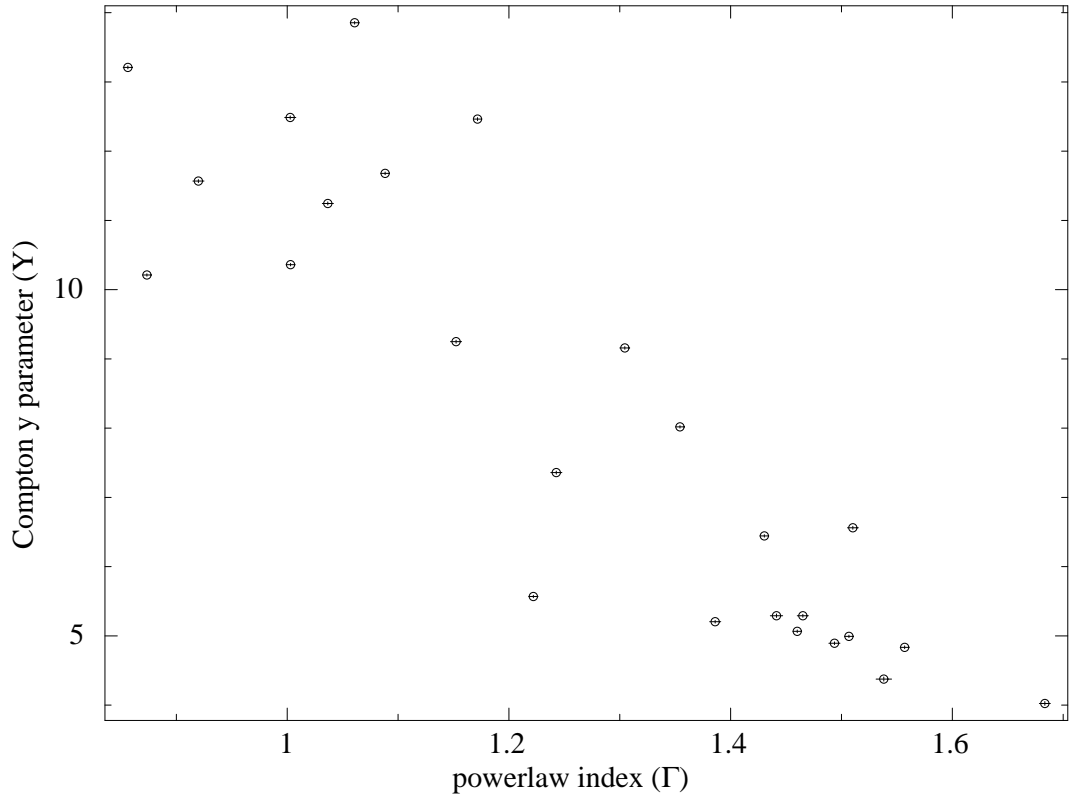}
\caption{Plot showing the interdependence between the power law index ($\Gamma$) and the Compton $Y$ parameter.
The error bars are plotted as seen within the symbols.}
\label{fig10}
\end{figure}

\begin{table*} 
\caption{Best fitting spectral parameters of Vela X-1 with a partial covering high-energy cutoff model. Errors quoted are for 99 per cent confidence range.}
\label{table1}
\centering
\begin{tabular}{@{}l c c@{} c}
\hline \hline
Parameter &  Highecut model values & CompTT model values\\
\hline
\ensuremath{N_{\mathrm{H1}}} ($10^{22}$ atoms $cm^{-2}$) & $1.38 \pm 0.01$ & $1.05_{-0.016}^{+0.017}$\\
\ensuremath{N_{\mathrm{H2}}} ($10^{22}$ atoms $cm^{-2}$)  & $4.00_{-0.26}^{+0.27}$ & $6.30_{-0.29}^{+0.30}$\\
\ensuremath{Cv_{\mathrm{fract}}} & $0.385 \pm 0.01 $ & $0.451 \pm 0.008$ \\
PowIndex &  $1.19 \pm 0.005$ & --\\
E-folding energy (keV) & $ 16.07_{-1.47}^{+1.66}$ & --\\
E-cut energy (keV) & $24.56_{-0.78}^{+0.80}$ & --\\
\ensuremath{powerlaw_{\mathrm{norm}}} $^a$ & $0.190 \pm 0.001$ & --\\
\ensuremath{CompTT_{\mathrm{T0}}} (keV) & -- &  $0.49 \pm 0.007$\\
CompTT $kT$ (keV) & -- & $ 10.59 \pm 0.14$ \\
CompTT $\tau$ & -- & $10.95 \pm 0.099$ \\
\ensuremath{CompTT_{\mathrm{norm}}} $^b$ & -- & $0.062 \pm 0.0004$ \\
\ensuremath{E1_{\mathrm{cycl}}} (keV) & $24.10_{-0.99}^{+1.05}$ & $24.42_{-1.57}^{+1.68} $\\
\ensuremath{D1_{\mathrm{cycl}}}  & $0.26 \pm 0.04$ & $0.28 \pm 0.05$\\
\ensuremath{W1_{\mathrm{cycl}}} (keV) & $4.37_{-1.39}^{+1.86} $ & $7.70_{-2.27}^{+2.97} $\\
\ensuremath{E2_{\mathrm{cycl}}} (keV) & $50.53_{-2.74}^{+3.36}$ & $48.70_{-1.23}^{+1.29}$\\
\ensuremath{D2_{\mathrm{cycl}}}  & $0.94_{-0.28}^{+0.33}$ & $1.82_{-0.17}^{+0.18}$\\
\ensuremath{W2_{\mathrm{cycl}}} (keV) & $10.00_{-3.28}^{+4.28} $  & $19.27_{-2.92}^{+3.80}$\\
Iron line energy (keV) & 6.41$\pm$ 0.01 & 6.41$\pm$ 0.01\\
Iron line eqwidth (\rmfamily{eV}) & 56.8 $\pm $2.3 & 56.7 $\pm $2.2\\
Flux (XIS) $^c$ (0.3-10 keV) &  $1.53 \pm 0.01$ & $1.53 \pm 0.01$\\
Flux (PIN) $^d$ (10-70 keV) &  $4.12 \pm 0.01$ & $4.11 \pm 0.01$\\
reduced $\chi^{2}$/d.o.f & 2.03/831 & 1.92/832\\
\hline
\end{tabular}\\
$^a$ \ensuremath{\mathrm{photons}\, \mathrm{keV}^{-1}\,\mathrm{cm}^{-2}\,\mathrm{s}^{-1}\,\mathrm{at}\, 1\, \mathrm{keV}} \\
$^b$ \ensuremath{\mathrm{photons}\, \mathrm{keV}^{-1}\,\mathrm{cm}^{-2}\,\mathrm{s}^{-1}\,\mathrm{at}\, 1\, \mathrm{keV}} \\
$^c$ Flux is in units of \ensuremath{10^{-9}\,  \mathrm{ergs}\,  \mathrm{cm}^{-2}\,\mathrm{s}^{-1}} and are in 99 \% confidence range.\\
$^d$ Flux is in units of  \ensuremath{10^{-9}\,  \mathrm{ergs}\,  \mathrm{cm}^{-2}\,\mathrm{s}^{-1}} and are in 99 \% confidence range.\\
\end{table*}

\end{document}